\documentclass[aps,12pt,showpacs,a4paper,floatfix,superscriptaddress]{revtex4}
\usepackage{longtable}
\usepackage{amsmath}
\usepackage{dcolumn}
\newcolumntype{f}[1]{D{.}{.}{#1}}
\ifx\pdftexversion\undefined
  \usepackage[dvips]{graphicx}
\else
  \usepackage[pdftex]{graphicx}
\fi

\begin{document}

\newcommand{\half}{\frac12}
\newcommand{\vare}{\varepsilon}
\newcommand{\eps}{\epsilon}
\newcommand{\pr}{^{\prime}}
\newcommand{\ppr}{^{\prime\prime}}
\newcommand{\pp}{{p^{\prime}}}
\newcommand{\hp}{\hat{\bfp}}
\newcommand{\hx}{\hat{\bfx}}
\newcommand{\hpp}{\hat{\bf p}\pr}
\newcommand{\hq}{\hat{\bfq}}
\newcommand{\rqq}{{\rm q}}
\newcommand{\rx}{{\rm x}}
\newcommand{\rp}{{\rm p}}
\newcommand{\rpp}{{{\rm p}^{\prime}}}
\newcommand{\rk}{{\rm k}}
\newcommand{\bfp}{{\bf p}}
\newcommand{\bfpp}{{\bf p}^{\prime}}
\newcommand{\bfq}{{\bf q}}
\newcommand{\bfx}{{\bf x}}
\newcommand{\bfk}{{\bf k}}
\newcommand{\bfz}{{\bf z}}
\newcommand{\bfr}{{\bf r}}
\newcommand{\bphi}{{\mbox{\boldmath$\phi$}}}
\newcommand{\balpha}{{\mbox{\boldmath$\alpha$}}}
\newcommand{\bsigma}{{\mbox{\boldmath$\sigma$}}}
\newcommand{\bomega}{{\mbox{\boldmath$\omega$}}}
\newcommand{\bmu}{{\mbox{\boldmath$\mu$}}}
\newcommand{\bvare}{{\mbox{\boldmath$\varepsilon$}}}
\newcommand{\bGamma}{{\mbox{\boldmath$\Gamma$}}}
\newcommand{\bTheta}{{\mbox{\boldmath$\Theta$}}}
\newcommand{\bLambda}{{\mbox{\boldmath$\Lambda$}}}
\newcommand{\bgamma}{{\mbox{\boldmath$\gamma$}}}
\newcommand{\bnabla}{{\mbox{\boldmath$\nabla$}}}
\newcommand{\bvarrho}{{\mbox{\boldmath$\varrho$}}}
\newcommand{\intzo}{\int_0^1}
\newcommand{\intinf}{\int^{\infty}_{-\infty}}
\newcommand{\ka}{\kappa_a}
\newcommand{\kb}{\kappa_b}
\newcommand{\lbr}{\langle}
\newcommand{\rbr}{\rangle}
\newcommand{\ThreeJ}[6]{
        \left(
        \begin{array}{ccc}
        #1  & #2  & #3 \\
        #4  & #5  & #6 \\
        \end{array}
        \right)
        }
\newcommand{\SixJ}[6]{
        \left\{
        \begin{array}{ccc}
        #1  & #2  & #3 \\
        #4  & #5  & #6 \\
        \end{array}
        \right\}
        }
\newcommand{\NineJ}[9]{
        \left\{
        \begin{array}{ccc}
        #1  & #2  & #3 \\
        #4  & #5  & #6 \\
        #7  & #8  & #9 \\
        \end{array}
        \right\}
        }
\newcommand{\Dmatrix}[4]{
        \left(
        \begin{array}{cc}
        #1  & #2   \\
        #3  & #4   \\
        \end{array}
        \right)
        }
\newcommand{\Dcase}[4]{
        \left\{
        \begin{array}{cl}
        #1  & #2   \\
        #3  & #4   \\
        \end{array}
        \right.
        }
\newcommand{\cross}[1]{#1\!\!\!/}

%
%
\title{
Evaluation of the self-energy correction to the $\boldsymbol{g}$-factor of
$\boldsymbol{S}$ states in H-like ions}
\author{V. A. Yerokhin}
\email{yerokhin@pcqnt1.phys.spbu.ru}
\affiliation{
 Department of Physics, St. Petersburg State University,
Oulianovskaya 1, Petrodvorets, St. Petersburg 198504, Russia}
\affiliation{
 Center for Advanced Studies, St. Petersburg State Polytechnical
University, Polytekhnicheskaya 29, St. Petersburg 195251, Russia}
\affiliation{
Laboratoire Kastler Brossel, \'Ecole Normale Sup\'erieure et
Universit\'e P. et M. Curie, Case 74, 4 place Jussieu, F-75252, Cedex 05,
France}
 \author{P. Indelicato}
\email{paul.indelicato@spectro.jussieu.fr}
\affiliation{
Laboratoire Kastler Brossel, \'Ecole Normale Sup\'erieure et
Universit\'e P. et M. Curie, Case 74, 4 place Jussieu, F-75252, Cedex 05,
France}
\author{V. M. Shabaev}
\affiliation{
 Department of Physics, St. Petersburg State University,
Oulianovskaya 1, Petrodvorets, St. Petersburg 198504, Russia}

\begin{abstract}
A detailed description of the numerical procedure is presented for the
evaluation of the one-loop self-energy correction to the $g$-factor of an
electron in the $1s$ and $2s$ states in H-like ions to all orders in
$Z\alpha$.
\end{abstract}

\maketitle

%
\section{Introduction}

Recently, several high-precision experiments have been performed on the
bound-electron $g$-factor in H-like carbon and oxygen by the Mainz-GSI
collaboration \cite{Haeffner00,Verdu02}. The value actually measured in the
experiment is $(g\, m/M)$, where $m$ is the electron mass, $M$ is the ion
mass, and $g$ is the $g$-factor of the electron. The relative accuracy of the
best experimental determination of this value \cite{Haeffner00} is $5\times
10^{-10}$, which is 4 times better than that of the accepted value for the
electron mass \cite{Mohr00}. Further progress is anticipated from the
experimental side, as well as extension of measurements to the higher-$Z$
region \cite{Werth01}.

The spectacular experimental results have triggered great interest to the
theoretical description of the $g$-factor of a bound electron
\cite{Blundell97,Persson97,Beier00,Czarnecki01,Karshenboim01,Karshenboim01a,
Shabaev01,Martynenko01,Glazov02,Beier02,Nefiodov02,Shabaev02PRL,Yerokhin02,Shabaev02,Yan02,Yerokhin02CJP,Shabaev03}.
Combining experimental values with accurate theoretical predictions for the
bound-electron $g$-factor resulted in an independent determination of the
electron mass \cite{Beier02,Yerokhin02}. The current accuracy of this
determination \cite{Yerokhin02} is 4 times better than that of the accepted
value for the electron mass \cite{Mohr00}. For the latest compilation of
various contributions to the bound-electron $g$-factor we refer the reader to
\cite{Yerokhin02CJP} for H-like ions and to \cite{Shabaev02,Shabaev03} for
Li-like ions.

In the present work we give a detailed description of our calculation of the
self-energy correction to the bound-electron $g$-factor for $1s$ and $2s$
states of a H-like ion. The first results of this calculation for the $1s$
state were previously published in \cite{Yerokhin02}, where they were used
for the determination of the electron mass. In this paper, we extend our
consideration to a higher-$Z$ region and perform calculations also for the
$2s$ state, having in mind the planned extension of the experiments to
Li-like systems.

Our calculation is carried out in the Feynman gauge. The relativistic units
($\hbar=c=1$) and the Heaviside charge units ($ \alpha = e^2/4\pi$, $e<0$)
are used throughout the paper. We also use the notations $\cross{p} =
p_{\mu}\gamma^{\mu}$ and $\hp = \bfp/|\bfp|$.

%
\section{Basic formulas}

In this paper we will consider a bound electron in an $s$ state of a H-like
ions with a spinless nucleus interacting with a static homogeneous magnetic
field. The bound-electron $g$-factor is defined by
\begin{equation}
g = - \frac{\lbr j_a m_a| \mu_z |j_am_a \rbr}{\mu_0\, m_a}\,,
\end{equation}
where $\bmu$ is the operator of the magnetic moment of the electron, $\mu_0 =
|e|/(2m)$ is the Bohr magneton, $j_a$ is the total angular momentum of the
electron, and $m_a$ is its projection. The lowest-order value for the $g$
factor can be found by a simple relativistic calculation based on the Dirac
equation \cite{Breit}. For an $s$ state and the point nucleus, it yields
\begin{equation}\label{eq4}
g_{D} = \frac23 \left(1+\frac{2\,\vare_a}{m}\right) \, ,
\end{equation}
where $\vare_a$ is the energy of the electron state.

\begin{figure}
\centering
\includegraphics[clip=true]{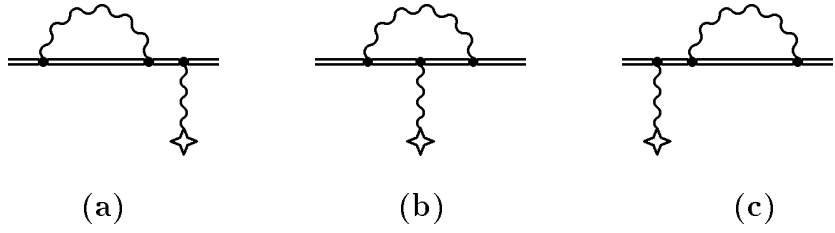}
\caption{Feynman diagrams representing the self-energy correction to the
bound-electron $g$-factor. The double line indicates the bound electron
propagator and the wave line that ends with a cross denotes the interaction
with the external magnetic field. \label{segfactor}}
\end{figure}

Various contributions to the $g$-factor are related to the corresponding
corrections to the energy shift, as given by
\begin{equation}    \label{eq2}
\Delta E = \Delta g\, \mu_0 B m_a\, ,
\end{equation}
where the magnetic field $\bf B$ is assumed to be directed along the $z$
axis. In the present investigation we are interested in the self-energy
correction to the bound-electron $g$-factor. It is diagrammatically depicted
in Fig.~\ref{segfactor}. Formal expressions for the corresponding
contributions can easily be derived, e.g., by the two-time Green function
method \cite{ShabaevReport}. They can be also obtained by considering the
first-order perturbation of the self-energy correction by a perturbing
magnetic potential $\delta V(\bfx)$ \cite{Indelicato91,Persson96}.

The contribution of the diagram (b) is referred to as the vertex ("ver") term
and is given by
\begin{eqnarray}    \label{eq5}
\Delta E_{\rm ver} &=&  \frac{i}{2\pi}\intinf d\omega\,
 \nonumber \\ && \times
    \sum_{n_1n_2}
    \frac{\lbr n_1|\delta V|n_2\rbr
            \lbr an_2| I(\omega)|n_1a\rbr}
                   {(\vare_a-\omega-u\,\vare_{n_1})(\vare_a-\omega-u\,\vare_{n_2})}\,,
\end{eqnarray}
where $\delta V(\bfx) = -e\,\balpha \cdot {\bf A}(\bfx)$, ${\bf A}$ is the
vector potential ${\bf A}(\bfx) = (1/2) [{\bf B}\times \bfx]$, $u = 1-i0$
where the small imaginary addition preserves the correct circumvention of
poles of the electron propagators, $I(\omega)$ is the operator of the
electron-electron interaction,
\begin{equation}
I(\omega) = e^2 \alpha^{\mu}\alpha^{\nu} D_{\mu\nu}(\omega)\,,
\end{equation}
${\alpha}^{\mu} = (1, \balpha) $ are the Dirac matrices, and $D_{\mu\nu}$ is
the photon propagator. The contribution of the diagrams (a) and (c) is
conveniently divided into 2 parts that are referred to as the irreducible and
the reducible one. The reducible ("red") contribution is defined as a part in
which the intermediate states in the spectral decomposition of the middle
electron propagator (between the self-energy loop and the magnetic
interaction) coincide with the initial valence state. The irreducible ("ir")
part is given by the remainder. It can be written in terms of non-diagonal
matrix elements of the self-energy function as
\begin{equation}    \label{eq6}
\Delta E_{\rm ir} = \lbr \delta a|\gamma^0 \widetilde{\Sigma}(\vare_a) |a\rbr
                  + \lbr a|\gamma^0 \widetilde{\Sigma}(\vare_a) |\delta a\rbr\,,
\end{equation}
where the perturbed wave function is given by
\begin{equation}    \label{eq7}
|\delta a\rbr = \sum_{n}^{\vare_n\ne\vare_a} \frac{|n\rbr \lbr n| \delta
        V|a\rbr}{\vare_a-\vare_n}\,,
\end{equation}
$\widetilde{\Sigma}(\vare) = \Sigma(\vare)-\delta m$, $\Sigma(\vare)$ is the
self-energy function defined as
\begin{eqnarray} \label{sefunction}
\Sigma(\vare,\bfx_1,\bfx_2) &=& 2i\alpha\gamma^0 \intinf d\omega\,
      \alpha_{\mu}
  \nonumber \\ && \times
         G(\vare-\omega,\bfx_1,\bfx_2) \alpha_{\nu}
 D^{\mu\nu}(\omega,x_{12})
                                 \,,
\end{eqnarray}
$\delta m$ is the mass counterterm, and $G$ is the Dirac-Coulomb Green
function, $G(\vare) = (\vare-H)^{-1}$, where $H$ is the Dirac-Coulomb
Hamiltonian. The expression for the reducible part reads
\begin{equation}    \label{eq8}
\Delta E_{\rm red} = \lbr a| \delta V| a\rbr \lbr a| \gamma^0 \left.
    \frac{\partial}{\partial \vare} \Sigma(\vare) \right|_{\vare = \vare_a} |
            a\rbr\,.
\end{equation}

%
\section{Detailed analysis}
\label{sec:det-an}

The irreducible contribution (\ref{eq6}) is written in terms of non-diagonal
matrix elements of the self-energy function. Its renormalization is well
known and we do not discuss it here. For the point nuclear model, the
perturbed wave function $|\delta a\rbr$ can be evaluated analytically in a
closed form by using the method of generalized virial relations for the Dirac
equation \cite{Shabaev91}. The explicit expression for the $|\delta a\rbr$
function can be found in \cite{Shabaev01}.

\begin{figure}
\centering
\includegraphics[clip=true]{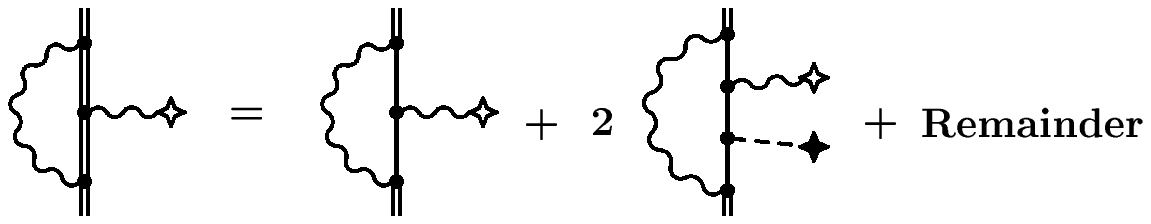}
\caption{The potential expansion of the vertex diagram.  The single line
indicates the free-electron propagator and the dashed line denotes the
interaction with the Coulomb field of the nucleus. The terms of the potential
expansion are referred to as the zero-potential, one-potential, and
many-potential contributions. \label{vertexRen}}
\end{figure}

Both the vertex and reducible contributions are ultraviolet (UV) divergent.
In order to covariantly renormalize them, we expand the bound-electron
propagators in Eqs. (\ref{eq5}) and (\ref{eq8}) in terms of the interaction
with the binding field. According to the number of interactions (with the
nuclear Coulomb field), the corresponding contributions are referred to as
the {\it zero-potential}, {\it one-potential}, and {\it many-potential}
terms. For the vertex diagram, this decomposition is schematically presented
in Fig.~\ref{vertexRen}.  We thus represent the vertex and reducible parts as
\begin{eqnarray} \label{eq9}
\Delta E_{\rm ver} = \Delta E^{(0)}_{\rm ver}+
                        \Delta E^{(1)}_{\rm ver}+
                        \Delta E^{(2+)}_{\rm ver} \,, \\
      \label{eq10}
\Delta E_{\rm red} = \Delta E^{(0)}_{\rm red}+
                        \Delta E^{(1)}_{\rm red}+
                        \Delta E^{(2+)}_{\rm red} \,,
\end{eqnarray}
where the superscript corresponds to the number of interactions with the
binding field. It is often convenient to consider the vertex and the
reducible terms together. In this case, we will use the following notation
\begin{equation}
\Delta E_{\rm vr}^{(i)} = \Delta E_{\rm ver}^{(i)}+ \Delta E_{\rm
red}^{(i)}\,,
\end{equation}
where $i = \left\{ 0, 1, 2+\right\}$.

By elementary power-counting arguments one can show that all terms containing
one or more interactions with the binding field are UV finite. Despite the
fact that the one-potential term is finite, we prefer to consider it
separately, as was first proposed in \cite{Persson97}. The reason for such
treatment is that, when evaluated in coordinate space, the one-potential term
yields a slowly-converging partial-wave expansion, which represents the main
computational difficulty in the low-$Z$ region. The numerical scheme
developed for the computation of this term in \cite{Persson97,Beier00}
allowed to extend the summation well beyond 100 partial waves, which explains
much better accuracy obtained in that work as compared to the first
evaluation \cite{Blundell97}. In our approach, we evaluate the one-potential
term directly in momentum space without utilizing the partial-wave expansion.
In this way we eliminate the uncertainty due to the estimation of the
uncalculated tail of the series.

%
\subsection{Zero-potential contribution}

The expression for the zero-potential vertex term can be obtained from
Eq.~(\ref{eq5}) by replacing all bound-electron propagators by the free
propagators. Writing it in momentum space, we have
\begin{equation}    \label{eq11}
\Delta E_{\rm ver}^{(0)} = -e \int \frac{d\bfp\, d\bfpp}{(2\pi)^6}\,
\overline{\psi}_a(\bfp)\, \bGamma_R(p,p\pr) \cdot {\bf A}(\bfp-\bfpp)\,
\psi_a(\bfpp)
 \,,
\end{equation}
where $\overline{\psi} = \psi^{\dag} \gamma^0$, $p$ and $p\pr$ are 4-vectors
with a fixed time component, $p = (\vare_a,\bfp)$, $p\pr = (\vare_a,\bfpp)$,
${\bf A}(\bfp-\bfpp)$ is the vector potential ${\bf A}(\bfx)$ in momentum
space,
\begin{equation}    \label{eq12}
{\bf A}(\bfp-\bfpp) = -\frac{i}2 (2\pi)^3 \left[{\bf B}\times \bnabla_{\bfpp}
\delta^3(\bfp-\bfpp)\right]
 \,,
\end{equation}
and $\Gamma^{\mu}_R (p,p\pr)$ is the UV-finite part of the free vertex
function introduced in Appendix~\ref{Aponeloop}. We note that right from the
beginning we are working with the renormalized expressions; the cancellation
of UV-divergent terms in the sum of the zero-potential vertex and reducible
terms can be checked explicitly.

Substituting Eq.~(\ref{eq12}) into Eq.~(\ref{eq11}) and separating the
contribution to the $g$-factor, we obtain
\begin{eqnarray}    \label{eq16}
\Delta g_{\rm ver}^{(0)} &=& -2im
    \int \frac{d\bfp\, d\bfpp}{(2\pi)^3}\, \overline{\psi}_a(\bfp)
\nonumber \\ && \times
    \left[\bnabla_{\bfpp} \delta^3(\bfp-\bfpp)
      \times \bGamma_R(p,p\pr) \right]_z \psi_a(\bfpp)\,,
\end{eqnarray}
where the gradient $\bnabla_{\bfpp}$ acts on the $\delta$ function only and
the angular-momentum projection of the initial state is assumed to be $m_a =
1/2$. The expression above is transformed by integrating by parts and
performing the integration that involves the $\delta$ function. The
integration by parts divides the whole contribution into two pieces
\begin{equation}    \label{eq17}
\Delta g_{\rm ver}^{(0)} = \Delta g_{\rm ver,1}^{(0)} + \Delta g_{\rm
ver,2}^{(0)} \,
\end{equation}
that correspond to the gradient acting on the vertex and the wave function,
respectively,
\begin{equation}    \label{eq18}
\Delta g_{\rm ver,1}^{(0)} = 2im \int \frac{d\bfp}{(2\pi)^3}\,
\overline{\psi}_a(\bfp)\, \Xi(p)\, \psi_a(\bfp)\,,
\end{equation}
\begin{equation}    \label{eq19}
\Delta g_{\rm ver,2}^{(0)} = -2im \int \frac{d\bfp}{(2\pi)^3}\,
\overline{\psi}_a(\bfp)  \left[ \bGamma_R(p,p) \times
    \bnabla_{\bfp} \right]_z \psi_a(\bfp) \,,
\end{equation}
where
\begin{equation}    \label{eq20}
\Xi(p) = \left. \left[ \bnabla_{\bfpp} \times
    \bGamma(p,p\pr) \right]_z \right|_{p\pr = p}\,.
\end{equation}

We start with the first contribution. The function $\Xi$ can be expressed as
\begin{eqnarray}    \label{eq21}
\Xi(p) &=& 4\pi i\alpha \int \frac{d^4k}{(2\pi)^4}
   \frac1{k^2} \gamma_{\sigma} \frac{\cross{p}-\cross{k}+m}
     {(p-k)^2-m^2}
\nonumber \\ && \times
     [\bgamma\times\bnabla_{\bfp}]_z
     \frac{\cross{p}-\cross{k}+m}{(p-k)^2-m^2}
       \gamma^{\sigma} \,.
\end{eqnarray}
Transformation of the numerator yields
\begin{eqnarray}    \label{eq22}
\Xi(p) &=& \frac{\alpha}{4\pi} \int \frac{d^4k}{i\pi^2} \frac1{k^2
    [(p-k)^2-m^2]^2 }
    \Bigl\{ \gamma_{\sigma} (\cross{p}-\cross{k}+m)
\nonumber \\      && \times
    [\bgamma\times\bgamma]_z
    \gamma^{\sigma}
    +2 \gamma_{\sigma} [\bgamma \times (\bfp-\bfk)]_z
    \gamma^{\sigma} \Bigr\} \,.
\end{eqnarray}
Basic integrals over the loop momentum $k$ can be easily evaluated to yield
\begin{eqnarray}    \label{eq23}
\int \frac{d^4k}{i\pi^2} \frac{\left\{1,k_{\mu}\right\}}
  {{k^2 [(p-k)^2-m^2}]^2 }
    &=& -\frac1{m^2} \int_0^1 dx\, \frac{\left\{1,x\,p_{\mu}\right\}}
    {(1-\rho)x+\rho}\,,
\nonumber \\
\end{eqnarray}
where $\rho = (m^2-p^2)/m^2$. Substituting these integrals into
Eq.~(\ref{eq22}) and taking into account that $[\bgamma\times\bgamma] = 2i
\gamma_0 \gamma_5 \bgamma$ and $\left\{\gamma_5,\gamma_{\mu}\right\} = 0$, we
obtain
\begin{equation}    \label{eq24}
\Xi(p) = \frac{\alpha}{\pi m^2} A(\rho) \biggl( i\gamma_5 \bgamma_z
\gamma_0\, \cross{p} - [\bgamma \times \bfp]_z \biggr) \,,
\end{equation}
where
\begin{equation}    \label{eq25}
A(\rho) = \frac1{1-\rho} \biggl( 1+\frac1{1-\rho} \ln \rho \biggr)\,.
\end{equation}

For $s$ states, the angular integration in Eq.~(\ref{eq18}) is carried out by
employing the following results for basic angular integrals ($\mu = 1/2$)
\begin{equation}    \label{eq26}
\int d\hp\, \chi^{\dag}_{\kappa \mu}(\hp)\, \sigma_z \chi_{\kappa \mu}(\hp)\,
=                      \Dcase{1\,,}{{\rm for}\ \kappa = -1\,,}
                            {-1/3\,,}{{\rm for}\ \kappa = 1\,,}
\end{equation}
\begin{equation}    \label{eq27}
\int d\hp\, \chi^{\dag}_{\kappa \mu}(\hp)\, [\bsigma \times \hp]_z
 \chi_{-\kappa \mu}(\hp)\, =
                      \Dcase{-2/3\, i\,,}{{\rm for}\ \kappa = -1\,,}
                            {2/3\, i\,,}{{\rm for}\ \kappa = 1\,,}
\end{equation}
where $\chi_{\kappa \mu}(\hp)$ is the spin-angular spinor \cite{Rose} and
$\bsigma$ denotes a vector consisting of the Pauli matrices. Finally, the
result for the first vertex contribution reads ($a$ is assumed to be an $s$
state)
\begin{eqnarray}    \label{eq28}
\Delta g_{\rm ver,1}^{(0)} &=& \frac{\alpha}{4\pi^4m} \int_0^{\infty}  dp_r\,
    p_r^2 A(\rho)
\nonumber \\ && \times
    \biggl\{ g_a(\vare_a g_a +p_r f_a)
    -\frac13 f_a (\vare_a f_a+
    p_r g_a)\biggr\} \ \,,
\end{eqnarray}
where $p_r = |\bfp|$, and $g_a = g_a(p_r)$ and $f_a = f_a(p_r)$ are the upper
and the lower components of the wave function, respectively.

Now we turn to the second vertex contribution given by Eq.~(\ref{eq19}). The
vertex function with two equal arguments $\bGamma_R(p,p)$ can be obtained by
the Ward identity,
\begin{equation}    \label{eq29}
\bGamma(p,p) = -\bnabla_{\bfp} \Sigma^{(0)}(p)\,,
\end{equation}
where $\Sigma^{(0)}(p)$ is the free self-energy function introduced in
Appendix \ref{Aponeloop}. Simple differentiation yields
\begin{eqnarray}    \label{eq30}
\bGamma_R(p,p) &=& \frac{\alpha}{4\pi} \biggl[ b_1(\rho)\, \bgamma +
b_2(\rho)\, \cross{p}\, \bfp + b_3(\rho)\, \bfp \biggr]\,,
   \\
b_1(\rho) &=& \frac{2-\rho}{1-\rho} \left( 1+ \frac{\rho}{1-\rho} \ln
\rho\right)
  \,, \\
b_2(\rho) &=& -\frac{2}{m^2} \frac1{(1-\rho)^2} \left( 3-\rho
  + \frac2{1-\rho} \ln \rho \right)\,,
  \\
b_3(\rho) &=& \frac8{m(1-\rho)} \left( 1+ \frac1{1-\rho}\ln \rho \right)
   \,.
\end{eqnarray}

In order to perform the integration over the angular variables in
Eq.~(\ref{eq19}), we use the representation for the gradient in the spherical
coordinates \cite{Varshalovich}
\begin{equation}    \label{eq34}
\bnabla_{\bfp} = \hp \frac{\partial}{\partial p_r} + \frac1{p_r}
\bnabla_{\Omega} \,.
\end{equation}
Now the angular integration can be carried out by using the following results
for basic angular integrals ($\mu = 1/2$)
\begin{equation}    \label{eq35}
\int d\hp\, \chi^{\dag}_{\kappa \mu}(\hp)\, [\hp \times \bnabla_{\Omega}]_z
 \chi_{\kappa \mu}(\hp)\, = \Dcase{0\,,}{{\rm for}\ \kappa = -1\,,}
                            {2/3\, i\,,}{{\rm for}\ \kappa = 1\,,}
\end{equation}
\begin{equation}    \label{eq36}
\int d\hp\, \chi^{\dag}_{\kappa \mu}(\hp)\, [\bsigma \times
\bnabla_{\Omega}]_z  \chi_{-\kappa \mu}(\hp)\, = \Dcase{\!-4/3\, i,}{{\rm
for}\,\kappa = -1\,,}
                            {0\,,}{{\rm for}\, \kappa = 1\,.}
\end{equation}
The final result for the second vertex contribution reads ($a$ is an $s$
state)
\begin{eqnarray}    \label{eq37}
\Delta g_{\rm ver,2}^{(0)} &=& -\frac{\alpha m}{24 \pi^4} \int_0^{\infty}
    dp_r\, p_r^2
\nonumber \\ && \times
    \Biggl[ b_1(\rho) \biggl( \frac2p_r g_af_a+ g_a \frac{df_a}{dp_r} -f_a \frac{dg_a}{dp_r}
    \biggr)
\nonumber \\ &&
    -b_2(\rho) (\vare_af_a+p_r g_a) f_a
    + b_3(\rho) f_a f_a \Biggr] \,.
\end{eqnarray}
This concludes our consideration of the zero-potential vertex contribution.

The described approach to the evaluation of the zero-potential vertex term
was first employed in \cite{Blundell97}. We mention also a different
treatment of this term suggested by the Swedish group
\cite{Persson97,Beier00}. In that work, the $\delta$ function in
Eq.~(\ref{eq12}) was approximated by a continuous Gaussian function with a
small cutoff parameter. An advantage of the presented approach is that we end
up with a single integration that remains to be performed numerically, while
the consideration of \cite{Persson97,Beier00} leaves a 4-dimensional
integration in final formulas. This complication is not crucial for the
zero-potential term since it is relatively simple. For the one-potential
term, however, the consideration is more difficult, and the approach based on
the integration by parts turns out to be extremely helpful.

The zero-potential reducible term is given by
\begin{equation}    \label{eq38}
\Delta g_{\rm red}^{(0)} = g_{D} \lbr a| \gamma^0 \left.
    \frac{\partial}{\partial \vare} \Sigma^{(0)}_R(\vare) \right|_{\vare = \vare_a} |
            a\rbr\,,
\end{equation}
where $g_{D}$ is the lowest-order value of the $g$-factor given by
Eq.~(\ref{eq4}). The derivative of the free self-energy function with respect
to the energy argument reads
\begin{eqnarray}  \label{eq39}
\left. \frac{\partial}{\partial\vare} \Sigma^{(0)}_R(p) \right|_{\vare =
\vare_a} &=& -\frac{\alpha}{4\pi}
      \biggl[ a_1(\rho)\, \cross{p}
      + a_2(\rho)\, \gamma_0 + a_3(\rho) \biggr] \, ,
\nonumber \\
\end{eqnarray}
where
\begin{eqnarray}
a_1(\rho) &=& -\frac{2\vare_a}{m^2 (1-\rho)^2}
        \left( 3-\rho +\frac2{1-\rho}\ln\rho \right) \,, \\
a_2(\rho) &=& 2+ \frac{\rho}{1-\rho}
        \left( 1+\frac{2-\rho}{1-\rho}\ln\rho \right) \, , \\
a_3(\rho) &=& \frac{8\vare_a}{m(1-\rho)}
        \left( 1 +\frac1{1-\rho}\ln\rho \right) \, .
\end{eqnarray}
Integration over the angular variables yields
\begin{eqnarray}    \label{eq40}
\Delta g_{\rm red}^{(0)} &=& g_{D} \left(-\frac{\alpha}{32\pi^4} \right)
        \int_0^{\infty} dp_r\, p_r^2
\nonumber \\ && \times
        \Biggl\{ a_1(\rho)\, \biggl[
        \vare_a(g_a^2+f_a^2)+2p_rg_af_a \biggr]
\nonumber \\ &&
  + a_2(\rho)\, \bigl( g_a^2+f_a^2\bigr)
    + a_3(\rho)\, \bigl(
        g_a^2-f_a^2\bigr) \Biggr\} \, .
\end{eqnarray}
Finally, the total zero-potential term $\Delta g_{\rm vr}^{(0)}$ is given by
the sum of Eqs.~(\ref{eq28}), (\ref{eq37}), and (\ref{eq40}).

%
%
%
\subsection{One-potential term}

The one-potential vertex contribution to the $g$-factor can be written as
\begin{eqnarray}    \label{eq41}
\Delta g^{(1)}_{\rm ver} &=& 4im \int \frac{d\bfp\, d\bfr\,
    d\bfpp}{(2\pi)^6}\,
   \overline{\psi}_a(\bfp)\, V_C(\bfr)
\nonumber \\ && \times
   \left[\bLambda(p,r,p\pr)
     \times \bnabla_{\bfr} \delta^3(\bfp-\bfpp-\bfr) \right]_z
       \psi_a(\bfpp) \,,
\end{eqnarray}
where the angular-momentum projection of the valence state is assumed to be
$m_a = 1/2$, $p = (\vare_a,\bfp)$,  $r = (\vare_a,\bfr)$, $p\pr =
(\vare_a,\bfpp)$, the function $\bLambda$ is defined by
\begin{eqnarray} \label{eq42}
\bLambda(p,r,p\pr) &=& -4\pi i\alpha \int \frac{d^4 k}{(2\pi)^4}
  \gamma_{\sigma} \frac{\cross{p}-\cross{k}+m}{(p-k)^2-m^2}
    \gamma_0
\nonumber \\ && \times
    \frac{\cross{p}-\cross{k}-\cross{r}+m}{(p-k-r)^2-m^2}
      \bgamma  \frac{\cross{p}\pr-\cross{k}+m}{(p\pr-k)^2-m^2}
         \gamma^{\sigma}\,,
\nonumber \\
\end{eqnarray}
and $V_C(\bfr) = -4\pi \alpha Z/\bfr^2$ is the Coulomb potential in momentum
space. Obtaining Eq.~(\ref{eq41}), we have employed the vector potential in a
form equivalent to Eq.~(\ref{eq12})
\begin{equation} \label{eq43}
{\bf A}(\bfp-\bfpp-\bfr) =  -\frac{i}2 (2\pi)^3 \left[{\bf B}\times
\bnabla_{\bfr} \delta^3(\bfp-\bfpp-\bfr)\right]\,.
\end{equation}
We mention also a factor of 2 included into Eq.~(\ref{eq41}) accounting for
two equivalent diagrams.

Integrating by parts, we divide the expression into two parts,
\begin{equation}    \label{eq44}
\Delta g^{(1)}_{\rm ver} = \Delta g^{(1)}_{\rm ver,1}+ \Delta g^{(1)}_{\rm
ver,2}\,,
\end{equation}
where
\begin{equation} \label{eq45}
\Delta g^{(1)}_{\rm ver,1} = 4im \int \frac{d\bfp\, d\bfpp}{(2\pi)^6}\,
   V_C(\bfq)\, \overline{\psi}_a(\bfp)  {\Theta}_z(p,p\pr)\,
       \psi_a(\bfpp) \,,
\end{equation}
\begin{eqnarray} \label{eq46}
\Delta g^{(1)}_{\rm ver,2} &=& -4im \int \frac{d\bfp\, d\bfpp}{(2\pi)^6}\,
   \overline{\psi}_a(\bfp)
\nonumber \\ && \times
   \left[ \bLambda(p,q,p\pr)
     \times {\bf S}(\bfq) \right]_z
       \psi_a(\bfpp) \,,
\end{eqnarray}
where $q = p-p\pr = (0,\bfp-\bfpp)$, ${\bf S}(\bfq) = \bnabla_{\bfq}
V_C(\bfq)$, and
\begin{equation}
{\bTheta}(p,p\pr) =  \left[
      \bnabla_{\bfr} \times   \bLambda(p,r,p\pr)
      \right]_{\bfr=\bfq}\,.
\end{equation}

\subsubsection{$\Delta g^{(1)}_{\rm ver,1}$ contribution}

We start our consideration of the $\Delta g^{(1)}_{\rm ver,1}$ term with the
function $\bTheta$,
\begin{eqnarray}
{\bTheta}(p,p\pr) &=&   4\pi i\alpha \int \frac{d^4 k}{(2\pi)^4}
  \gamma_{\sigma} \frac{\cross{p}-\cross{k}+m}{(p-k)^2-m^2}
    \gamma_0
\nonumber \\ && \times
    \left[ \bnabla_{\bfpp}
    \frac{\cross{p}\pr-\cross{k}+m}{(p\pr-k)^2-m^2} \times
      \bgamma \right]
    \frac{\cross{p}\pr-\cross{k}+m}{(p\pr-k)^2-m^2}
         \gamma^{\sigma}\,,
\nonumber \\
\end{eqnarray}
where the gradient acts on the expression in the square brackets only. By
using the following identity
\begin{equation}
\bnabla_{p\pr}
    \frac{\cross{p}\pr-\cross{k}+m}{(p\pr-k)^2-m^2} =
\frac{\cross{p}\pr-\cross{k}+m}{(p\pr-k)^2-m^2} \bgamma
\frac{\cross{p}\pr-\cross{k}+m}{(p\pr-k)^2-m^2}\,,
\end{equation}
and commutation relations for the $\gamma$ matrices, we obtain
\begin{eqnarray}
{\bTheta}(p,p\pr) &=&   -2\pi i\alpha \int \frac{d^4 k}{(2\pi)^4}
 \nonumber \\ && \times
    \frac{ \bvarrho}{
     k^2 \bigl[(p-k)^2-m^2 \bigr] \bigl[(p\pr-k)^2-m^2 \bigr]^2
     } \,,
\end{eqnarray}
where the numerator $\bvarrho$ is
\begin{eqnarray}
\bvarrho &=& \gamma_{\sigma} (\cross{p}-\cross{k}+m) \gamma_0 \Bigl\{
     (\cross{p}\pr-\cross{k}+m) \left[ \bgamma \times \bgamma \right]
\nonumber \\ &&
     + \left[ \bgamma \times \bgamma \right] (\cross{p}\pr-\cross{k}+m)
     \Bigr\} \gamma^{\sigma}\,.
\end{eqnarray}

Now we employ the identity $\left[ \bgamma \times \bgamma \right] = 2i
\gamma_0 \gamma_5 \bgamma$ and use the commutation relations in order to
bring the matrices $\gamma_0$, $\gamma_5$, $\bgamma$ to the right. This
yields
\begin{eqnarray}
\bvarrho &=& 4i \gamma_{\sigma} (\cross{p}-\cross{k}+m)
 \nonumber \\ && \times
   \Bigl[ m\bgamma
   + (p\pr_0-k_0)\gamma_0\bgamma -(\bfpp-\bfk) \Bigr]
   \gamma^{\sigma} \gamma_5 \,.
\end{eqnarray}
Carrying out the summation over the repeating indices in the numerator, we
obtain
\begin{eqnarray}
\bvarrho &=&  8i
   \gamma_5 \Bigl[ m^2 \bgamma +2m\bfq -(\bfpp-\bfk)(\cross{p}-\cross{k})
\nonumber \\ &&
  +(p\pr_0-k_0)\bgamma \gamma_0 (\cross{p}-\cross{k}) \Bigr]\,.
\end{eqnarray}

The integration over the loop momentum $k$ is carried out by using the
following results for basic integrals
\begin{eqnarray}
& \displaystyle    \int \frac{d^4 k}{i\pi^2}
       \frac{ \left\{ 1,k_{\mu},k_{\mu}k_{\nu}\right\}
     }{
     k^2 \bigl[(p-k)^2-m^2 \bigr] \bigl[(p\pr-k)^2-m^2 \bigr]^2
     } =
\nonumber \\ & \displaystyle
     \int_0^1 dx\,dy\, \frac{1-y}{N^2} \left\{ 1, x
        b_{\mu}, x^2 b_{\mu}b_{\nu}
       -\frac{xN}{2} g_{\mu\nu} \right\} \,,
\end{eqnarray}
where $b = yp+(1-y)p\pr$, $N = xb^2 +y m^2 \rho + (1-y)m^2 \rho\pr$, $\rho\pr
= (m^2-{p\pr}^2)/m^2$.

By an explicit calculation one can show that the part of the basic integrals
proportional to $g_{\mu\nu}$ yields a vanishing contribution. We therefore
have
\begin{eqnarray}
{\bTheta}(p,p\pr) &=&   \frac{i\alpha}{\pi}
       \int_0^1 dx\,dy\, \frac{1-y}{N^2} \gamma_5
\nonumber \\ && \times
       \Bigl[
       m^2 \bgamma + 2m\bfq -{\bf Q}\pr \cross{Q} +
Q\pr_0 \bgamma \gamma_0 \cross{Q}\Bigr] \,,
\end{eqnarray}
where $Q = p-xb = (1-xy)p-x(1-y)p\pr$, $Q\pr = p\pr-xb = -xyp +(1-x+xy)p\pr$.
We now use the commutation relations in order to bring $\cross{p}$ to the
left of $\bgamma$ and $\cross{p}\pr$ to the right of $\bgamma$. This yields
\begin{eqnarray}
{\bTheta}(p,p\pr) &=&   \frac{i\alpha}{\pi}
       \int_0^1 dx\,dy\, \frac{1-y}{N^2} \gamma_5
       \Bigl\{A_0\bgamma +(C_1\cross{p}+C_2\cross{p}\pr)\bfp
\nonumber \\ &&
         +(D_1\cross{p}+D_2\cross{p}\pr)\bfpp
    + F_1\, \cross{p}\gamma_0 \bgamma + F_2\, \bgamma \gamma_0 \cross{p}\pr
\nonumber \\ &&
  +    G_1\bfp + G_2\bfp\pr+ H_1\bfp\gamma_0 \Bigr\}\,,
\end{eqnarray}
where the coefficient functions are given by
\begin{eqnarray}
 A_0 &=& m^2+ 2 \vare_a^2(1-x)(1-xy)\,, \nonumber \\
 C_1 &=& xy(1-xy) \,,\nonumber  \\
 C_2 &=& -x^2y(1-y) \,,\nonumber \\
 D_1 &=& -(1-x+xy)(1-xy)\,, \nonumber \\
 D_2 &=&  (1-x+xy)x(1-y)\,, \\
 F_1 &=& -\vare_a (1-x)(1-xy)\,, \nonumber \\
 F_2 &=& -\vare_a (1-x)x(1-y) \,,\nonumber \\
 G_1 &=& 2m \,,\nonumber \\
 G_2 &=& -2m \,,\nonumber \\
 H_1 &=& -2\vare_a(1-x)(1-xy) \,. \nonumber
\end{eqnarray}


In order to perform the angular integration in Eq.~(\ref{eq45}), we define
the set of scalar functions ${\cal P}_i$ ($i = 1\ldots 6$):
\begin{eqnarray}
\overline{\psi}_a(\bfp) {\bTheta}(p,p\pr) \psi_a(\bfpp) &=&
-\frac{i\alpha}{\pi} \int_0^1 dx\, dy\,
    \Biggl\{
{\cal P}_1\, \chi^{\dag}_{+}(\hp)\bsigma\chi_{+}(\hpp) + {\cal P}_2\,
\chi^{\dag}_{-}(\hp)\bsigma\chi_{-}(\hpp)
   \nonumber \\
&& + \bfp\, {\cal P}_3\, \chi^{\dag}_{+}(\hp)\chi_{-}(\hpp)
   + \bfp\, {\cal P}_4\, \chi^{\dag}_{-}(\hp)\chi_{+}(\hpp)
   \nonumber \\
&&   + \bfpp\, {\cal P}_5\, \chi^{\dag}_{+}(\hp)\chi_{-}(\hpp)
   + \bfpp\, {\cal P}_6\, \chi^{\dag}_{-}(\hp)\chi_{+}(\hpp)
   \Biggr\} \,,
\end{eqnarray}
where $\chi_{\pm}(\hp) = \chi_{\pm \kappa_a, m_a}(\hp)$. The functions ${\cal
P}_{i}$ depend on  $p_r = |\bfp|$, $p^{\prime}_r = |\bfp^{\prime}|$, and
$\xi=\hp\cdot\hpp$ only. They are given by
\begin{eqnarray}
{\cal P}_1 &=& \frac{1-y}{N^2} \biggl[ A_0g_ag_a\pr +F_1(\vare_a g_a+p_r
f_a)g_a\pr
     + F_2 g_a(\vare_a g_a\pr +p\pr_r f_a\pr) \biggr] \,, \nonumber\\
{\cal P}_2 &=& \frac{1-y}{N^2} \biggl[ A_0f_af_a\pr +F_1(\vare_a f_a+p_r
g_a)f_a\pr
     + F_2 f_a(\vare_a f_a\pr +p\pr_r g_a\pr) \biggr] \,, \nonumber\\
{\cal P}_3 &=& \frac{1-y}{N^2} \biggl[ C_1(\vare_a g_a+p_r f_a)f_a\pr +C_2
g_a(\vare_a f_a\pr+p\pr_r g_a\pr)
     + (H_1-G_1)g_af_a\pr \biggr] \,, \nonumber\\
{\cal P}_4 &=& \frac{1-y}{N^2} \biggl[ C_1(\vare_a f_a+p_r g_a)g_a\pr +C_2
f_a(\vare_a g_a\pr+p\pr_r f_a\pr)
     + (H_1+G_1)f_ag_a\pr \biggr] \,, \\
{\cal P}_5 &=& \frac{1-y}{N^2} \biggl[ D_1(\vare_a g_a+p_r f_a)f_a\pr +D_2
g_a(\vare_a f_a\pr+p\pr_r g_a\pr)
     -G_2 g_af_a\pr \biggr] \,, \nonumber\\
{\cal P}_6 &=& \frac{1-y}{N^2} \biggl[ D_1(\vare_a f_a+p_r g_a)g_a\pr +D_2
f_a(\vare_a g_a\pr+p\pr_r f_a\pr)
     +G_2 f_ag_a\pr \biggr] \,,\nonumber
\end{eqnarray}
where $g_a = g_a(p_r)$, $f_a = f_a(p_r)$, $g_a\pr = g_a(p\pr_r)$, $f_a\pr =
f_a(p\pr_r)$ are the radial components of the valence wave function.

Now the angular integration in Eq.~(\ref{eq45}) can be performed, taking into
account the following results for basic angular integrals ($\mu = 1/2$)
\begin{eqnarray}
\int d\hp\, d\hpp\, F\, \chi^{\dag}_{\kappa \mu}(\hp)\,
                 \bsigma_z\, \chi_{\kappa \mu}(\hpp)
 &=& \Dcase{2\pi \displaystyle \int_{-1}^1 d\xi\, F\,,}{\mbox{ for $\ \kappa = -1$}\,}
 {-2\pi/3 \displaystyle \int_{-1}^1 d\xi\, \xi F\,,}{\mbox{ for $\ \kappa = 1$}\,} \\
\int d\hp\, d\hpp\, F\, \chi^{\dag}_{\kappa \mu}(\hp)\,
                  \hp_z \,\chi_{-\kappa \mu}(\hpp)
 &=& \Dcase{-2\pi/3 \displaystyle \int_{-1}^1 d\xi\, \xi F\,,}{\mbox{for $\ \kappa = -1$}\,,}
 {-2\pi/3 \displaystyle \int_{-1}^1 d\xi\,  F\,,}{ \mbox{for $\ \kappa = 1$}\,,} \\
\int d\hp\, d\hpp\, F\, \chi^{\dag}_{\kappa \mu}(\hp)\,
                  \hpp_z \,\chi_{-\kappa \mu}(\hpp)
 &=& \Dcase{-2\pi/3 \displaystyle \int_{-1}^1 d\xi\, F\,,}{\mbox{for $\ \kappa = -1$}\,,}
 {-2\pi/3 \displaystyle \int_{-1}^1 d\xi\, \xi F\,,}{\mbox{for $\ \kappa =
 1$}\,,}
\end{eqnarray}
where $F$ is a function that depends on $p_r$, $p\pr_r$, and $\xi$ only.
Finally, we obtain the expression for the first vertex contribution,
\begin{eqnarray}   \label{eqone1}
\Delta g^{(1)}_{\rm ver,1} &=& m\frac{\alpha^2 Z}{6\pi^5} \int_0^{\infty}
dp_r\, dp_r\pr\, \int_{-1}^{1} d\xi\, \int_0^1 dx\, dy\,
    \frac{p_r^2\, {p_r\pr}^2}{q_r^2}
\nonumber \\  && \times
    \Bigl[-3 {\cal P}_1+ \xi {\cal P}_2+
       p_r(\xi {\cal P}_3+ {\cal P}_4)+ p_r\pr({\cal P}_5+ \xi {\cal P}_6) \Bigr]\,.
\end{eqnarray}

\subsubsection{$\Delta g^{(1)}_{\rm ver,2}$ contribution}

First, we evaluate the function $\bLambda$ in Eq.~(\ref{eq46}). Its
expression can be obtained from the vertex function by the following identity
\begin{equation}
\bLambda(p,q,p\pr) =  \bnabla_{\bfpp} \Gamma^0(p,p\pr)\,,
\end{equation}
where $\Gamma^{\mu}(p,p\pr)$  is defined Appendix \ref{Aponeloop}. Using
Eqs.~(\ref{v3})-(\ref{v5a}) and taking into account that
\begin{equation}
\bnabla_{\bfpp} N = 2(1-y)[-xy\, \bfp + (1-x+xy)\bfpp]\,,
\end{equation}
\begin{equation}
\bnabla_{\bfpp} a = -x \bnabla_{\bfpp} N + 2(1-x)\bfp \,,
\end{equation}
we obtain
\begin{eqnarray}
\bLambda(p,q,p\pr) &=&
   -\frac{\alpha}{2\pi} \int_0^1 dx\, dy\, \Biggl\{ \frac{2(1-y)}{N^2}
   \biggl( -xy\,\bfp+(1-x+xy)\bfpp \biggr)
\nonumber \\ && \times
   \biggl[ (a+m^2+2xN)\gamma_0 +\vare_a B \cross{p} +\vare_a C \cross{p}\pr
    + D \cross{p}\gamma_0 \cross{p}\pr + \vare_a H \biggr]
   \nonumber \\ &&
+ \frac1N \biggl( -2(1-x)\bfp \gamma_0 + \vare_a C\bgamma + D \cross{p}
\gamma_0
  \bgamma \biggr) \Biggr\} \,,
\end{eqnarray}
with the coefficient functions $a$ and $B$--$D$ given by
Eqs.~(\ref{v4})-(\ref{v5}). Note that the expression in the square brackets
is similar to the corresponding term in Eq.~(\ref{v3}). We can therefore
write
\begin{eqnarray}
\overline{\psi}_a(\bfp) \biggl[  (a+m^2+2xN)\gamma_0 +\vare_a B \cross{p}
+\vare_a C \cross{p}\pr
    + D \cross{p}\gamma_0 \cross{p}\pr + \vare_a H \biggr]  \psi_a(\bfpp) =
\nonumber \\
     \hat{{\cal F}}_1\,
\chi^{\dag}_{+}(\hp)\chi_{+}(\hpp) + \hat{{\cal F}}_2\,
\chi^{\dag}_{-}(\hp)\chi_{-}(\hpp) \,,
\end{eqnarray}
where the expressions for $\hat{{\cal F}}_{1,2}$ can be obtained from
(\ref{v7}), (\ref{v8}) by a straightforward substitution. Finally, we obtain
\begin{eqnarray}
\overline{\psi}_a(\bfp)\, \bLambda(p,q,p\pr)\, \psi_a(\bfpp) &=&
-\frac{\alpha}{2\pi} \int_0^1 dx\, dy\,
    \Biggl\{
{\cal R}_1\, \chi^{\dag}_{+}(\hp)\,\bsigma\,\chi_{-}(\hpp) + {\cal R}_2\,
\chi^{\dag}_{-}(\hp)\,\bsigma\,\chi_{+}(\hpp)\,
   \nonumber \\
&& + \bigl(\bfp {\cal R}_3+\bfpp {\cal R}_4 \bigr)
\chi^{\dag}_{+}(\hp)\,\chi_{+}(\hpp)
 + \bigl(\bfp {\cal R}_5+\bfpp {\cal R}_6 \bigr) \chi^{\dag}_{-}(\hp)\,\chi_{-}(\hpp)\
   \Biggr\} \,, \nonumber \\
\end{eqnarray}
where
\begin{eqnarray}
{\cal R}_1 &=& \frac1N \bigl[ \vare_a Cg_af_a\pr +D(\vare_a g_a+p_r
    f_a)f_a\pr \bigr] \,, \nonumber \\
 {\cal R}_2 &=& \frac1N \bigl[ \vare_a Cf_ag_a\pr +D(\vare_a f_a+p_r g_a)g_a\pr
    \bigr] \,, \nonumber \\
 {\cal R}_3 &=& \frac{2(1-y)}{N^2}(-xy) \hat{{\cal F}}_1 -\frac{2(1-x)}{N}
    g_ag_a\pr \,, \nonumber \\
 {\cal R}_4 &=& \frac{2(1-y)}{N^2}(1-x+xy) \hat{{\cal F}}_1
    \,,\\
 {\cal R}_5 &=& \frac{2(1-y)}{N^2}(-xy) \hat{{\cal F}}_2 -\frac{2(1-x)}{N}
    f_af_a\pr \,, \nonumber \\
 {\cal R}_6 &=& \frac{2(1-y)}{N^2}(1-x+xy) \hat{{\cal F}}_2
        \,, \nonumber
\end{eqnarray}
and
\begin{eqnarray}
\hat{{\cal F}}_1 &=&
  (a+m^2+2xN+\vare_a H) g_ag_a\pr
  \nonumber \\ &&
  +\vare_a B(\vare_a g_a+p_r f_a)g_a\pr + \vare_a C g_a(\vare_a g_a\pr+p\pr_r f_a\pr)
  \nonumber \\ &&
            + D(\vare_a g_a+p_r f_a)(\vare_a g_a\pr+p\pr_r f_a\pr) \,, \\
\hat{{\cal F}}_2 &=&
  (a+m^2+2xN-\vare_a H)f_af_a\pr
  \nonumber \\ &&
    +\vare_a B(\vare_a f_a+p_r g_a)f_a\pr + \vare_a C f_a(\vare_a
        f_a\pr+p\pr_r g_a\pr)
  \nonumber \\ &&
            + D(\vare_a f_a+p_r g_a)(\vare_a f_a\pr+p\pr_r g_a\pr) \,.
\end{eqnarray}

We perform the angular integration in Eq.~(\ref{eq46}) taking into account
the identity
\begin{equation}
\bnabla_{\bfq} V_C(\bfq) = 8\pi \alpha Z \frac{\hq}{q_r^3}\,,
\end{equation}
and  the following basic angular integrals ($\mu = 1/2$)
\begin{eqnarray}
\int d\hp\, d\hpp\, F\, \chi^{\dag}_{\kappa \mu}(\hp)\,
               [\hq \times \bsigma]_z \,\chi_{-\kappa \mu}(\hpp)
 &=& \Dcase{\displaystyle -4\pi i/3 \int_{-1}^1 d\xi\, \frac{(p_r\pr-\xi
              p_r)}{q_r}F\,,}{ \mbox{for $\ \kappa = -1$}\,,}
              {\displaystyle -4\pi i/3 \int_{-1}^1 d\xi\, \frac{(p_r-\xi
              p_r\pr)}{q_r}F\,,}{ \mbox{for $\ \kappa = 1$}\,,}\\
\int d\hp\, d\hpp\, F\, \chi^{\dag}_{\kappa \mu}(\hp)\,
               [\hp \times \hpp]_z \,\chi_{\kappa \mu}(\hpp)
 &=& \Dcase{0\,,}{\mbox{for $\ \kappa = -1$}\,,}
 {2\pi i/3 \displaystyle \int_{-1}^1 d\xi\, (1-\xi^2)F\,,}
                                               {\mbox{for $\ \kappa = 1$}\,,}
\end{eqnarray}
where $F$ is a function that depends on $p_r$, $p\pr_r$, and $\xi$ only. The
final result for the second vertex contribution reads
\begin{eqnarray}   \label{eqone2}
\Delta g^{(1)}_{\rm ver,2} &=& -m\frac{\alpha^2 Z}{3\pi^5} \int_0^{\infty}
dp_r\, dp_r\pr\, \int_{-1}^1 d\xi\,
     \int_0^1 dx\, dy\,
\nonumber \\ && \times
     \frac{p_r^2\, {p_r\pr}^2}{q_r^3}
       \Biggl[\frac{p\pr_r-\xi p_r}{q_r} {\cal R}_1+
            \frac{p_r-\xi p\pr_r}{q_r}{\cal R}_2 -\frac{(1-\xi^2)p_rp\pr_r}{2q_r} \bigl({\cal R}_5
      +{\cal R}_6 \bigr) \Biggr]\,.
\end{eqnarray}


\subsubsection{Reducible part}

The one-potential reducible contribution is given by
\begin{eqnarray}   \label{eqone3}
\Delta g_{\rm red}^{(1)} &=& g_{D} \int \frac{d\bfp\, d\bfpp}{(2\pi)^6}\,
   V_C(\bfq)\,
\nonumber \\ && \times
 \overline{\psi}_a(\bfp)\,
   \frac{\partial}{\partial \vare_a}
       \Gamma^{0}(p,p\pr)\, \psi_a(\bfpp) \,,
\end{eqnarray}
where $\Gamma^{\mu}(p,p\pr)$ is the vertex function, $p = (\vare_a,\bfp)$,
$p\pr = (\vare_a,\bfpp)$, and $g_{D}$ is the Dirac value of the $g$-factor.
An expression for this contribution can be obtained by a simple
differentiation of formulas for $\Gamma^{\mu}(p,p\pr)$ given in
Appendix~\ref{Aponeloop}.

%
\subsection{Many-potential term}

Expressions for the many-potential term can be obtained by the point-by-point
subtraction of the corresponding zero- and one-potential contributions from
the unrenormalized expressions (\ref{eq5}) and (\ref{eq8}). In order to
transform them to the form suitable for a numerical evaluation, we have to
perform the summation over the magnetic substates and the integration over
the angular variables. This can be carried out by using standard techniques
(see, e.g., \cite{Yerokhin99}). We present here only the results for the
unrenormalized contributions. The final expressions can be obtained from them
by the corresponding point-by-point subtractions.

The vertex correction to the $g$-factor is given by ($a$ is an $s$ state)
\begin{eqnarray}   \label{manyver}
\Delta g_{\rm ver} &=& \frac{i\alpha}{2\pi} \intinf d\omega\,
    \sum_{n_1n_2J} \SixJ{1/2}{1/2}{1}{j_2}{j_1}{J}
\nonumber \\ && \times
    \frac{ P(n_1,n_2)\, R_J(\omega,an_2n_1a)}
    {(\vare_a-\omega-u\vare_{n_1})(\vare_a-\omega-u\vare_{n_2})}\,,
\end{eqnarray}
where $u = 1-i0$, $R_J$ is the generalized Slater integral \cite{Johnson88a}
(the corresponding expressions can be found in \cite{Yerokhin99}),
\begin{eqnarray}
P(n_1,n_2) &=& \Pi(l_1l_20) (-1)^{l_1} \sqrt{\frac{(2j_1+1)(2j_2+1)}{3}}
 \nonumber \\ && \times
    \ThreeJ{j_2}{j_1}{1}{1/2}{1/2}{-1} R_{n_1n_2} \,,
\end{eqnarray}
$\Pi(l_1l_2l_3) = [1+(-1)^{l_1+l_2+l_3}]/2$, $(\cdots)$ and $\{ \cdots \}$
denote $3j$ and $6j$ symbols, respectively, and
\begin{equation}
R_{n_1n_2} = \int_0^{\infty} dr\, r^3
            \bigl[g_{n_1}(r)f_{n_2}(r)+f_{n_1}(r)g_{n_2}(r)\bigr]\,.
\end{equation}
The reducible contribution reads ($a$ is an $s$ state)
\begin{eqnarray}   \label{manyred}
\Delta g_{\rm red} &=& -\frac{i\alpha}{4\pi} \intinf d\omega\,
 \nonumber \\ && \times
     \sum_{nJ}
    \frac{ (-1)^{j_a-j_n+J} P(a,a)\,
    R_J(\omega,anna)}{(\vare_a-\omega-u\vare_n)^2}\,.
\end{eqnarray}

We mention that both the vertex and the reducible contribution are infrared
(IR) divergent. The divergence occurs when $\vare_{n_1} = \vare_{n_2} =
\vare_a$ in the vertex term and $\vare_n  =  \vare_a$ in the reducible term.
However, the sum of these terms can be shown to be convergent. In practical
calculations, we perform the $\omega$ integration for the sum of the vertex
and the reducible contributions. In that case, the integrand is a regular
function for small values of $\omega$ and can be numerically integrated up to
the desirable accuracy.

%
\section{High-precision evaluation of the irreducible contribution}
\label{sec:irred}

In this section we describe adaptations of the approach developed in
 \cite{Indelicato92,Indelicato98,Indelicato01}. In this method, the
 renormalization scheme is completely performed in
 coordinate space, which provide a great simplification in the numerical evaluation. In that method the integration over
 the photon energy $z$ is done last. The complex plane integration contour is divided in two parts \cite{Mohr74}.
 The contour around the bound state poles of the electron propagator gives the low-energy part, which reads, in Feynman gauge:
 \begin{equation}
\Delta  E_{\textrm{ir,L}} =  \frac{-2\alpha} {\pi}{\rm
P}\!\int_0^{\varepsilon_a} d z \int d\boldsymbol{x}_2 \int d\boldsymbol{x}_1
\ \psi_a^\dagger (\boldsymbol{x}_2) \alpha^{\mu}
G(\boldsymbol{x}_2,\boldsymbol{x}_1, z) \alpha_{\mu} \delta
a(\boldsymbol{x}_1) \frac{\sin [(\varepsilon_a-z)x_{21}] }{ x_{21}} \ ,
\label{eq:low_edwf}
\end{equation}
where $\delta a$ is the first order wave function correction \eqref{eq7}, $\varepsilon_a$ is the energy and $\psi_a$ the wavefunction
of the bound state under consideration
and P denotes a principal value integral.
 Here we use the Feynman gauge for the low-energy part instead of the Coulomb gauge in Ref.~\cite{Indelicato01}
 to be compatible with the gauge used in the vertex and reducible part calculations. Here the integrand is well enough behaved
 that the transformation of the low-energy part in Coulomb gauge done in \cite{Mohr74} to improve numerical accuracy
 is no longer required. The high energy parts is written as
 \begin{eqnarray}
\Delta  E_{\textrm{ir,H}}&=&\frac{-\alpha}{ \pi {\rm i}} \int_{\rm C_H} d z\ \int d\boldsymbol{x}_2 \int
d\boldsymbol{x}_1 \, \psi_a^\dagger (\boldsymbol{x}_2) \alpha_\mu
{G}(\boldsymbol{x}_2,\boldsymbol{x}_1, z)
 \alpha^\mu \delta a(\boldsymbol{x}_1) \frac{e^{-bx_{21}}}{ x_{21}} \nonumber \\
  && -2\delta m\int
  d\boldsymbol{x}\ \psi_a^\dagger (\boldsymbol{x}) \beta \delta a(\boldsymbol{x}) \ ,
\label{eq:high_e}
\end{eqnarray}
 \noindent where $b=-i\left[({\varepsilon_a}-z)^2
+i\delta\right]^{1/2}, {\rm Re}(b)>0$, and $\boldsymbol{x}_{21} = \boldsymbol{x}_2 -
\boldsymbol{x}_1$.  The index $\mu$ is summed from
0 to 3.  The contour ${\rm C_H}$ extends from $-i\infty$ to $0-i\epsilon$
and from $0+i\epsilon$ to $+i\infty$, with the appropriate branch of
$b$ chosen in each case. Note that the sign change in \eqref{eq:low_edwf} and \eqref{eq:high_e} compared to Refs.~\cite{Mohr74,Indelicato01} comes from
the opposite convention $G(\varepsilon)=(H-\varepsilon)^{-1}$ used in those works.

In this part we use the Pauli-Villars regularization and
\begin{equation}
\delta m({\mit \Lambda})=\frac{\alpha}{\pi}\left[\frac{3}{4}
\ln({\mit \Lambda}^2)+\frac{3}{8}\right] \ .
\end{equation}
Because we do not do a transformation to Coulomb gauge in \eqref{eq:low_edwf} angular integration is now identical to the one done for the high-energy part
\eqref{eq:high_e} as described in \cite{Mohr74}, and lead to the same angular coefficients.

The high-energy part is split into two pieces $\Delta  E_{\textrm{ir,HA}}$ and
$\Delta  E_{\textrm{ir,HB}}$. In $\Delta  E_{\textrm{ir,HA}}$
 the Coulomb Green's function is expanded around the free one up to the one
potential term, the wavefunction is expanded around $\boldsymbol{x}_1\approx
\boldsymbol{x}_ 2$, which is the place from which singularities that lead to
ultraviolet divergence arise. This part must thus contains the regularization
term. The regularized integral is evaluated analytically. The high-energy
remainder $\Delta  E_{\textrm{ir,HB}}$, which is  finite is evaluated
numerically. The details of the method are described
 in Refs.~\cite{Indelicato92,Indelicato98,Indelicato01}. Because the
cancellations that occurs in the normal self-energy (the contribution is
 formally of order $\alpha m$, while the final result is of order $\alpha m
(Z\alpha)^{4}$, leading to the loss of 9 significant
 figures at $Z=1$), are not present here, very high accuracy
 can be reached even at low-$Z$. At $Z=1$ one looses only 3.5 significant
figures for $1s$ state and 4 for $2s$ states. Convergence of the numerical
evaluation of the integrals is checked by doing a sequence of numerical
 calculations with increasing number of integration points in all three
integrations. In the case of $Z=1$, we have observed that the value obtained
with this method, provides a final answer in slight disagreement
 with the results from the $Z\alpha$ expansion, even though the convergence of
the numerical integration would lead to think that accuracy is larger than the
observed difference. This loss of accuracy is probably due
 to the  code used for the numerical evaluation of the Green function in the
high-energy part $\Delta  E_{\textrm{ir,HB}}$. Due to that fact, we do not use
the method described for this section for $Z=1$, and for
 $Z=2$, we used the half-sum of the value obtained with the method of this
section and from the one described in Sec.~\ref{sec:det-an}. The different
contributions to the irreducible part, evaluated by the method
 described in this section are presented in tables \ref{tab:irr-conf} and
\ref{tab:irr-conf-2s} for $1s$ and $2s$ respectively.
 \begin{table}
 \caption{Contributions to the $1s$ irreducible correction in unit of $10^{-6}$ (ppm). Note the cancellation between the low energy part and the piece of the
 high-energy part containing the renormalization terms. Numbers in parenthesis represent error in the last digit (when not given, accuracy is better than
  1 on the last digit)\label{tab:irr-conf}}
\begin{ruledtabular}
 \squeezetable
\begin{tabular}{cf{14}f{11}f{11}f{9}f{10}}
\multicolumn{1}{c}{$Z$} &
 \multicolumn{1}{c}{$\Delta  g_{\textrm{ir,L}}$} &
 \multicolumn{1}{c}{$\Delta  g_{\textrm{ir,HA}}$} &
 \multicolumn{1}{c}{$\Delta  g_{\textrm{ir,L}}+\Delta  g_{\textrm{ir,HA}}$} &
 \multicolumn{1}{c}{$\Delta  g_{\textrm{ir,HB}}$} &
 \multicolumn{1}{c}{$\Delta g_{\textrm{ir}}$} \\
 \hline
2 & -5414.138671 (1) & 5419.00973735696 & 4.871066 & 0.3353273 (5) & 5.206394 (1) \\
3 & -5408.112191 (1) & 5417.88129395137 & 9.769103 & 0.7540304 (8) & 10.523133 (2) \\
4 & -5400.4246855 (6) & 5416.30141200512 & 15.8767265 & 1.3394077 (10) & 17.216134 (1) \\
5 & -5391.25332317 (6) & 5414.27003034210 & 23.01670717 & 2.090731 (1) & 25.107438 (1) \\
6 & -5380.72953754 (3) & 5411.78707032559 & 31.05753279 & 3.007139 (1) & 34.064671 (1) \\
8 & -5356.015609415 (3) & 5405.46601341913 & 49.450404004 & 5.331304 (2) & 54.781708 (2) \\
10 & -5326.8970673996 (3) & 5397.33726330666 & 70.4401959071 & 8.303600 (2) & 78.743796 (2) \\
12 & -5293.8034464388 (6) & 5387.39956343566 & 93.5961169969 & 11.915577 (2) & 105.511694 (2) \\
13 & -5275.86794401 (1) & 5381.75188865495 & 105.88394464 & 13.958983 (3) & 119.842927 (3) \\
14 & -5257.05051270478 (1) & 5375.65137934454 & 118.60086663976 & 16.159514 (3) & 134.760380 (3) \\
15 & -5237.37900574235 (5) & 5369.09780041185 & 131.71879466950 & 18.516447 (3) & 150.235242 (3) \\
16 & -5216.87799726641 (5) & 5362.09089951271 & 145.21290224630 & 21.029202 (3) & 166.242105 (3) \\
18 & -5173.4722361824 (4) & 5346.71603642683 & 173.2438002444 & 26.520679 (3) & 199.764480 (3) \\
20 & -5126.98086779847 (5) & 5329.52442786062 & 202.54356006215 & 32.632888 (4) & 235.176448 (4) \\
24 & -5025.19605057045 (1) & 5289.68016155027 & 264.48411097982 & 46.733467 (4) & 311.217578 (4) \\
30 & -4851.76112169811 & 5216.21351982533 & 364.45239812722 & 72.745514 (5) & 437.197912 (5) \\
32 & -4788.66975035492 & 5188.05668759197 & 399.38693723705 & 82.795193 (5) & 482.182130 (5) \\
40 & -4511.24490648394 (1) & 5056.97546181288 & 545.73055532894 & 130.860760 (6) & 676.591315 (6) \\
50 & -4111.58003121901 (6) & 4851.22331387135 & 739.64328265234 & 213.227113 (7) & 952.870396 (7) \\
54 & -3936.57321146902 & 4755.74050579472 & 819.16729432570 & 255.361384 (8) & 1074.528678 (8) \\
60 & -3659.38321787930 (6) & 4598.23534473287 & 938.85212685357 & 331.513253 (8) & 1270.365380 (8) \\
70 & -3163.4138601297 (1) & 4297.20241652793 & 1133.7885563982 & 504.469815 (9) & 1638.258372 (9) \\
80 & -2635.3139826863 (2) & 3947.32337723397 & 1312.0093945477 & 760.56376 (1) & 2072.57316 (1) \\
82 & -2527.08776432844 & 3871.42617585973 & 1344.33841153129 & 825.21671 (1) & 2169.55512 (1) \\
83 & -2472.7536638438 (2) & 3832.73421755273 & 1359.9805537089 & 859.54725 (1) & 2219.52780 (1) \\
90 & -2089.95631351091 (3) & 3547.99624047955 & 1458.03992696864 & 1143.17384 (1) & 2601.21377 (1) \\
92 & -1980.36334087885 (9) & 3462.17838439369 & 1481.81504351484 & 1240.35520 (1) & 2722.17025 (1) \\
\end{tabular}
\end{ruledtabular}
\end{table}

 \begin{table}
 \caption{Contributions to the $2s$ irreducible correction in unit of $10^{-6}$ (ppm). Note the cancellation between the low energy part and the piece of the
 high-energy part containing the renormalization terms. Numbers in parenthesis represent error in the last digit (when not given, accuracy is better than
  1 on the last digit)\label{tab:irr-conf-2s}}
\begin{ruledtabular}
 \squeezetable
\begin{tabular}{cf{14}f{11}f{11}f{9}f{10}}
\multicolumn{1}{c}{$Z$} &
 \multicolumn{1}{c}{$\Delta  g_{\textrm{ir,L}}$} &
 \multicolumn{1}{c}{$\Delta  g_{\textrm{ir,HA}}$} &
 \multicolumn{1}{c}{$\Delta  g_{\textrm{ir,L}}+\Delta  g_{\textrm{ir,HA}}$} &
 \multicolumn{1}{c}{$\Delta  g_{\textrm{ir,HB}}$} &
 \multicolumn{1}{c}{$\Delta g_{\textrm{ir}}$} \\
 \hline
2 & -5418.28010 (7) & 5419.68677266709 & 1.40667 & 0.0838500 (3) & 1.49053 (7) \\
3 & -5416.53887 (4) & 5419.40461685172 & 2.86575 & 0.1886438 (4) & 3.05439 (4) \\
4 & -5414.29096 (2) & 5419.00952526959 & 4.71857 & 0.3352848 (6) & 5.05385 (2) \\
5 & -5411.582306 (5) & 5418.50142439885 & 6.919118 & 0.5237213 (7) & 7.442839 (5) \\
6 & -5408.447839 (5) & 5417.88021960316 & 9.432381 & 0.7538938 (8) & 10.186274 (5) \\
8 & -5401.0080206 (7) & 5416.29801360386 & 15.2899930 & 1.339196 (1) & 16.629189 (1) \\
10 & -5392.143636 (1) & 5414.26172425270 & 22.118088 & 2.090740 (1) & 24.208828 (2) \\
12 & -5381.981776 (2) & 5411.76982341076 & 29.788047 & 3.008216 (1) & 32.796263 (2) \\
13 & -5376.446112 (1) & 5410.35244545649 & 33.906333 & 3.529151 (2) & 37.435484 (2) \\
14 & -5370.6216706 (6) & 5408.82043238619 & 38.1987618 & 4.091577 (2) & 42.290338 (2) \\
15 & -5364.5179797 (4) & 5407.17349348969 & 42.6555138 & 4.695544 (2) & 47.351057 (2) \\
16 & -5358.143743 (2) & 5405.41131523977 & 47.267572 & 5.341130 (2) & 52.608702 (2) \\
18 & -5344.614996 (3) & 5401.53987070156 & 56.924875 & 6.757632 (2) & 63.682507 (4) \\
20 & -5330.092484 (3) & 5397.20312270768 & 67.110639 & 8.342381 (2) & 75.453020 (4) \\
24 & -5298.258097 (1) & 5387.11987055704 & 88.861774 & 12.024994 (3) & 100.886767 (3) \\
30 & -5244.1212783 & 5368.40791014636 & 124.2866318 & 18.883088 (3) & 143.169720 (3) \\
32 & -5224.5050123 & 5361.19426002437 & 136.6892477 & 21.544762 (3) & 158.234010 (3) \\
40 & -5138.91407892 (6) & 5327.29401168822 & 188.37993277 & 34.271932 (4) & 222.651865 (4) \\
50 & -5017.61128816 (1) & 5272.96663863740 & 255.35535048 & 55.742549 (5) & 311.097900 (5) \\
54 & -4965.162351525 & 5247.25973707291 & 282.097385548 & 66.481454 (5) & 348.578840 (5) \\
60 & -4882.7432641428 (7) & 5204.11518737168 & 321.3719232289 & 85.466426 (5) & 406.838349 (5) \\
70 & -4736.49467585151 & 5118.83522473171 & 382.34054888020 & 126.900793 (6) & 509.241342 (6) \\
80 & -4580.59370963553 (1) & 5014.30902799881 & 433.71531836328 & 185.322546 (7) & 619.037864 (7) \\
82 & -4548.34480741164 & 4990.74964795914 & 442.40484054750 & 199.688616 (7) & 642.093456 (7) \\
83 & -4532.08744880436 (1) & 4978.60543848221 & 446.51798967785 & 207.267271 (7) & 653.785261 (7) \\
90 & -4415.68500839307 (1) & 4886.18969971863 & 470.50469132556 & 268.797768 (8) & 739.302459 (8) \\
92 & -4381.51458962614 (4) & 4857.17690140483 & 475.66231177869 & 289.515129 (8) & 765.177441 (8) \\\end{tabular}
\end{ruledtabular}
\end{table}

%
\section{Numerical evaluation and results}

We start with reporting some details about our numerical evaluation. The
calculation for $1s$ and $2s$ states was performed for the point nucleus. In
addition, for the ground state, we repeated our evaluation for the
hollow-shell model of the nuclear-charge distribution and tabulated the
difference as the nuclear-size effect $\Delta g_{\rm NS}$.

The calculation of the irreducible part is quite straightforward. For the
point nucleus, the perturbed wave function $|\delta a \rbr$ can be found
analytically in a closed form \cite{Shabaev01}. In calculations involving an
extended nucleus, the perturbed wave function was evaluated numerically,
stored on a grid and then obtained in an arbitrary point by interpolation.
The calculation of a non-diagonal matrix element of the self-energy function
was carried out by two independent numerical methods. The first one is based
on an expansion of the bound-electron propagator in terms of interactions
with the binding field \cite{Snyderman91} (for details of the numerical
procedure see \cite{Yerokhin99}), whereas the second one is described in
Sec.~\ref{sec:irred}. Numerical results obtained by both methods are in good
agreement with each other. Point-nucleus results for the irreducible part are
listed in Tables \ref{table1s} and \ref{table2s}. Presented values are
obtained the by the second method in all cases, except $Z=1$ and 2 for the
$1s$ state. For $Z=1$, we used the result obtained by the first method; for
$Z=2$, a half-sum of the values obtained by the two methods was employed.

The evaluation of the zero-potential and one-potential contributions $\Delta
g^{(0)}_{\rm vr}$ and $\Delta g^{(1)}_{\rm vr}$ is relatively simple. The
zero-potential contribution [given by Eqs. (\ref{eq28}), (\ref{eq37}), and
(\ref{eq40})] contains a single numerical integration that can easily be
carried out to arbitrary precision. The one-potential contribution [Eqs.
(\ref{eqone1}), (\ref{eqone2}), and (\ref{eqone3})] contains a 5-dimensional
integration. The integration over one of the Feynman parameters can be
carried out analytically, which speeds up the calculation significantly. The
evaluation of this contribution is rather similar to that of the
one-potential term to the first-order self-energy correction
\cite{Yerokhin99}. Point-nucleus results for the zero-potential and the
one-potential contribution are listed correspondingly in the third and the
forth column of Tables \ref{table1s} and \ref{table2s}.

The evaluation of the many-potential contribution is the most difficult
numerical part of the present investigation and mainly defines the total
uncertainty of the results. Since the actual calculation of the
many-potential term is carried out by taking the point-by-point difference of
the unrenormalized, the free, and the one-potential contributions, one should
be aware about large numerical cancellations that occur in this difference.
Additional cancellations appear when the vertex term is added to the
reducible contribution. The final formulas for the many-potential term
contain an infinite summation over the angular-momentum parameter. (This
expansion is often referred to as the {\it partial-wave} expansion.) In our
approach, we chose the absolute value of the relativistic angular parameter
$\kappa$ of intermediate electron states, $|\kappa| = j+1/2$, to be the
expansion parameter. The summation was evaluated up to $|\kappa_{\rm max}| =
25-35$, and the tail of the expansion was estimated by a least-squares
inverse-polynomial fitting.

The general scheme of the evaluation of the many-potential contribution is
similar to that of our previous investigation \cite{Yerokhin01a}. A new
feature introduced in this work in the case of the $2s$ state is that we do
not introduce subtractions anymore that remove infrared divergences
separately in the vertex and the reducible term. Instead, we evaluate the
integration over $\omega$ for the {\it sum} of Eqs. (\ref{manyver}) and
(\ref{manyred}). In this case, infrared singularities that are present in two
parts of the integrand cancel each other and the total integrand is a smooth
regular function. This modification of the numerical procedure allows us to
avoid an additional loss of accuracy due to numerical cancellations.

The following contour was used for the $\omega$ integration in the present
work: $(\vare_0-i\infty, \vare_0-i0] + [\vare_0-i0,-i0] +[i0,\vare_0+i0]
+[\vare_0+i0,\vare_0+i\infty)$. This contour is advantageous for the
evaluation of self-energy corrections in the low-$Z$ region since it avoids
the appearance of pole terms that lead to additional numerical cancellations.
The parameter $\vare_0$ in the definition of the contour can be varied. In
actual calculations its value was taken to be about $Z\alpha\, \vare_a$ for
low $Z$.

The results of our numerical evaluation for the $1s$ and $2s$ states are
presented in Tables \ref{table1s} and \ref{table2s}, respectively. For the
ground state, we list both the point-nucleus and the extended-nucleus result.
It should be noted that the uncertainty specified for the latter result
refers to the estimated numerical error only and does not include the
nuclear-model dependence. We estimate the model dependence of the
nuclear-size effect $\Delta g_{\rm NS}$ to be about 1\%. Numerical values of
the root-mean-square radii used in our evaluation coincide with those of
\cite{Beier00}. For the ground state, we compare our values with the results
of the previous evaluations \cite{Blundell97,Beier00}. The results of
\cite{Blundell97} were obtained for the point nucleus, while the evaluation
\cite{Beier00} was carried out employing the homogeneously-charged spherical
model for the nuclear-charge distribution.

\begin{figure}
\centering
\includegraphics[clip=true,width=0.7\textwidth]{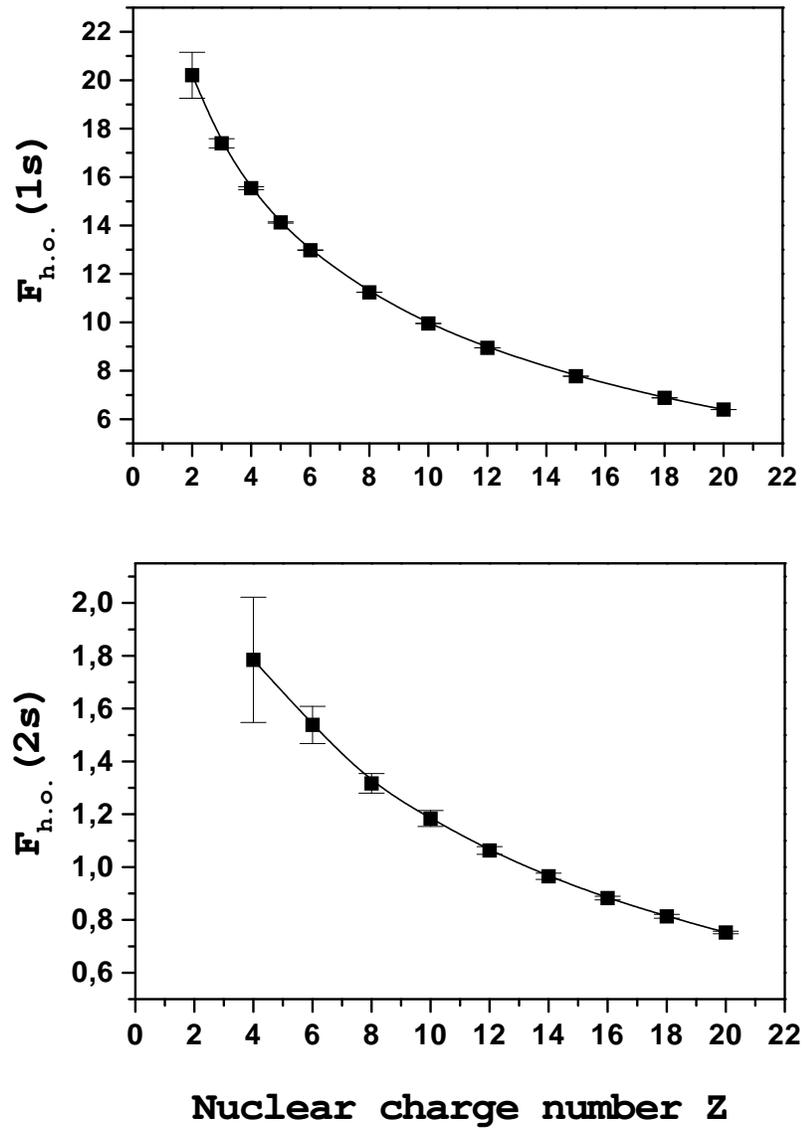}
\caption{ The higher-order self-energy contribution $F_{\rm h.o.}(Z\alpha)$
for the $1s$ and $2s$ electron $g$-factors in H-like ions. \label{fhofigure}}
\end{figure}

Finally, we compare our numerical values with the analytical results based on
the $Z\alpha$ expansion and isolate the higher-order contribution $F_{\rm
h.o.}(Z\alpha)$ that incorporates terms of order $(Z\alpha)^4$ and higher,
\begin{equation}
\Delta g_{\rm SE} = \frac{\alpha}{\pi} \left[ 1 +  (Z\alpha)^2\,a_{20} +
(Z\alpha)^4
   F_{\rm h.o.}(Z\alpha) \right]\,,
\end{equation}
where the first term in the brackets is the known Schwinger correction, the
second term $a_{20}$ was derived first by Grotch \cite{Grotch70} for the $1s$
state and later generalized to $ns$ states by Shabaev {\it et al.}
\cite{Shabaev02},
\begin{equation}
a_{20} = \frac1{6n^2}\,.
\end{equation}
The higher-order function $F_{\rm h.o.}(Z\alpha)$ for $1s$ and $2s$ states is
plotted in Fig.~\ref{fhofigure}.

%
\section{Conclusion}

In this paper we have presented our evaluation of the one-loop self-energy
correction to the electron $g$-factor of $1s$ and $2s$ states in H-like ions.
As compared to the previous calculations of this correction for the $1s$
state, an improvement of accuracy of about an order of magnitude has been
achieved in the low-$Z$ region. For the most interesting experimental cases,
H-like carbon and oxygen, our calculation improved the accuracy of the
theoretical prediction for the $g$-factor by a factor of 3 for carbon and by
a factor of 2 for oxygen \cite{Yerokhin02}, which reduced the uncertainty of
the electron-mass determination based on these values. The new value for the
electron mass is \cite{Haeffner00,Yerokhin02}
\begin{equation}
m_e = 0.000\, 548\, 579\, 909\, 29\, (29)(8)\,,
\end{equation}
where the first uncertainty originates from the experimental value for the
ratio of the electronic Larmor precession frequency and the cyclotron
frequency of the ion in the trap, and the second error comes from the
theoretical value for the bound-electron $g$-factor.

%
\section*{Acknowledgements}

This study was supported in part by RFBR (Grant No. 01-02-17248), by Ministry
of Education (Grant No. E02-3.1-49), and by the program "Russian
Universities" (Grant No. UR.01.01.072). V.Y. acknowledges the support from
the Minist\`ere de l'Education Nationale et de la Recherche, the foundation
"Dynasty", and International Center for Fundamental Physics. The computation
was partly performed on the CINES and IDRIS French national computer centers.
Laboratoire Kastler Brossel is Unit{\'e} Mixte de Recherche du CNRS
n$^{\circ}$ 8552.

%
%
%
\appendix

\section{Free one-loop functions}
\label{Aponeloop}

The free self-energy function in the Feynman gauge and in $D$ dimensions is
given by
\begin{equation}  \label{fse1}
\Sigma^{(0)}(p) = -4\pi i \alpha \mu^{2\eps} \int \frac{d^D k}{(2\pi)^D}
   \frac1{k^2}\gamma_{\sigma} \frac{\cross{p}-\cross{k}+m}{(p-k)^2-m^2}
     \gamma^{\sigma}\,.
\end{equation}
UV divergences in the above expression are regularized by working in $D =
4-2\epsilon$ dimensions. The mass parameter $\mu$ is introduced in order to
keep the proper dimension of the interaction term in the Lagrangian. We
separate the UV-finite part of the self-energy function $\Sigma_R^{(0)}$ as
follows
\begin{equation}    \label{fse4}
\Sigma^{(0)}(p) = \delta m-\frac{\alpha C_{\eps}}{4\pi\eps} (\cross{p}-m)
+\Sigma^{(0)}_{R}(p) \,,
\end{equation}
where the mass counterterm is given by
\begin{equation}    \label{fse3}
\delta m =  \frac{\alpha C_{\eps}}{4\pi\eps}\,
        \frac{3-2\eps}{1-2\eps}\, m \,,
\end{equation}
and
\begin{equation} \label{eq15}
C_{\eps} = \Gamma(1+\eps)(4\pi)^{\eps}
\left(\frac{\mu^2}{m^2}\right)^{\eps}\,.
\end{equation}
In the limit $\epsilon \to 0$, the renormalized part of the self-energy
function is given by
\begin{eqnarray}        \label{fse7}
\Sigma^{(0)}_{R}(p) &=& \frac{\alpha}{4\pi} \Biggl[ 2m \left( 1
+\frac{2\rho}{1-\rho}
    \ln \rho \right)
      \nonumber \\ &&
 - \cross{p}\, \frac{2-\rho}{1-\rho} \left( 1+\frac{\rho}{1-\rho} \ln \rho \right)
        \Biggr]\,,
\end{eqnarray}
where $\rho = (m^2-p^2)/m^2$.

The free vertex function in the Feynman gauge and in $D$ dimensions is
written as
\begin{eqnarray}     \label{eq13}
\Gamma^{\mu}(p,\pp) &=& -4\pi i \alpha\, \mu^{2\eps} \int
\frac{d^Dk}{(2\pi)^D}
    \frac1{k^2} \gamma_{\sigma}
\nonumber \\ && \times
    \frac{\cross{p}-\cross{k}+m}{(p-k)^2-m^2}
    \gamma^{\mu} \frac{\cross{p}^{\pr}-\cross{k}+m}{(\pp-k)^2-m^2}
    \gamma^{\sigma} \, .
\end{eqnarray}
The divergent part of the vertex function can be separated in the form
\begin{equation}    \label{eq14}
\Gamma^{\mu}(p,p\pr) = \frac{\alpha C_{\eps}}{4\pi \eps} \gamma^{\mu} +
\Gamma^{\mu}_{R}(p,p\pr)\,.
\end{equation}
Here we present an explicit expression only for the time component of the
renormalized vertex function $\Gamma^{\mu}_{R}$ omitting terms of order
$\epsilon$ and higher and assuming that $p^0 = {p\pr}^0 = \vare_a$,
\begin{eqnarray} \label{v3}
\Gamma^0_R(p,p\pr) &=& \frac{\alpha}{2\pi} \int_0^1 dx\, dy\,
    \frac1{N} \Bigl[ A\gamma_0
\nonumber \\ &&
    + \vare_a B \cross{p} + \vare_a C \cross{p}\pr
    + D \cross{p} \gamma_0 \cross{p}\pr + \vare_a H \Bigr]\,,
\end{eqnarray}
where
\begin{eqnarray} \label{v4}
 A &=& a+ m^2 -N\left( \frac34 + x\ln N \right)\,, \nonumber \\
 B &=& 2(1-xy)(1-x)\,,\nonumber \\
 C &=& 2(1-x+xy)(1-x) \,, \\
 D &=& -(1-x)\,,\nonumber \\
 H &=& -4m(1-x) \,,\nonumber
\end{eqnarray}
and
\begin{eqnarray} \label{v5}
a &=& xy(1-xy)p^2+x(1-y)(1-x+xy){p\pr}^2
 \nonumber \\ &&
 -2(1-xy)(1-x+xy)(p\cdot p\pr)\,,
\end{eqnarray}
\begin{equation} \label{v5a}
N = x \left[yp+(1-y)p\pr \right]^2 +y m^2 \rho + (1-y)m^2 \rho\pr\,.
\end{equation}
The integration over one of the Feynman parameters in Eq.~(\ref{v3}) can
easily be carried out leading to an expression, equivalent to that in
\cite{Yerokhin99}. However, we prefer to keep the vertex function in a more
compact form (\ref{v3}) here.

For carrying out angular integrations, we introduce the scalar functions
${\cal F}_{1,2}$ that depend on $p_r = |\bfp|$, $p_r^{\prime} =
|\bfp^{\prime}|$, and $\xi=\hp\cdot\hpp$ only,
\begin{eqnarray} \label{v6}
& \displaystyle \overline{\psi}_a\, (\bfp) \Gamma^0_{R}(p,p\pr)\,
\psi_a(\bfpp)
=
    \frac{\alpha}{2\pi} \int_0^1 dx\, dy\,
        \frac1{N}
\nonumber \\ & \displaystyle \times
    \biggl[
{\cal F}_1\, \chi^{\dag}_{+}(\hp)\,\chi_{+}(\hpp) + {\cal F}_2\,
\chi^{\dag}_{-}(\hp)\,\chi_{-}(\hpp)
  \biggr]\,,
\end{eqnarray}
\begin{eqnarray} \label{v7}
{\cal F}_1 &=& (A+\vare_a H) g_ag_a\pr +\vare_a B(\vare_a g_a+p_rf_a)g_a\pr
\nonumber \\ &&
  + \vare_a C g_a(\vare_a g_a\pr+p_r\pr f_a\pr)
\nonumber \\ &&
   + D(\vare_a g_a+p_rf_a)(\vare_a g_a\pr+p_r\pr f_a\pr) \,, \\
   \label{v8}
{\cal F}_2 &=& (A-\vare_a H)f_af_a\pr +\vare_a B(\vare_a f_a+p_rg_a)f_a\pr
    \nonumber \\ &&
+\vare_a C f_a(\vare_a f_a\pr+p_r\pr g_a\pr)
 \nonumber \\ &&
           + D(\vare_a f_a+p_rg_a)(\vare_a f_a\pr+p_r\pr g_a\pr) \,,
\end{eqnarray}
where $\chi_{\pm}(\hp) = \chi_{\pm \kappa_a, m_a}(\hp)$,  and $g_a = g_a(p)$,
$f_a = f_a(p)$, $g_a\pr = g_a(p\pr)$, and $f_a\pr = f_a(p\pr)$ are the radial
components of the valence wave function.

%

%
%
\newpage
\setlength{\LTcapwidth}{\textwidth}
\begingroup
\begin{longtable*}{rr@{}lr@{}lr@{}lr@{}lr@{}lr@{}lr@{}l}
\caption{ The one-loop self-energy correction to the $1s$-electron $g$-factor
for H-like ions. All values are absolute contributions to the $g$-factor
($1/\alpha = 137.035\,989\,5$) and presented in units of $10^{-6}$ (ppm).
Individual contributions are listed for the point nuclear model. $\Delta
g_{\rm NS}$ denotes the nuclear-size correction calculated for the shell
model of the nuclear-charge distribution. The labels "pnt." and "ext." refer
to the point-nucleus and the extended-nucleus result, respectively.
\label{table1s}}\\
\hline \hline $Z$ & \multicolumn{2}{c}{$\Delta g_{\rm ir}$} &
                    \multicolumn{2}{c}{$\Delta g_{\rm vr}^{(0)}$} &
                                    \multicolumn{2}{c}{$\Delta g_{\rm vr}^{(1)}$}
                                                   & \multicolumn{2}{c}{$\Delta g_{\rm vr}^{(2+)}$}
                                                                &\multicolumn{2}{c}{ $\Delta g_{\rm SE}$ (pnt.)}
                                                                                   &\multicolumn{2}{c}{$\Delta g_{\rm NS}$ }
                                                                                                    & \multicolumn{2}{c}{$\Delta g_{\rm SE}$ (ext.)} \\
\colrule
\endfirsthead
\caption{$1s$ $g$-factor. (continued)}\\
\hline \hline $Z$ & \multicolumn{2}{c}{$\Delta g_{\rm ir}$} &
                    \multicolumn{2}{c}{$\Delta g_{\rm vr}^{(0)}$} &
                                    \multicolumn{2}{c}{$\Delta g_{\rm vr}^{(1)}$}
                                                   & \multicolumn{2}{c}{$\Delta g_{\rm vr}^{(2+)}$}
                                                                &\multicolumn{2}{c}{ $\Delta g_{\rm SE}$ (pnt.)}
                                                                                   &\multicolumn{2}{c}{$\Delta g_{\rm NS}$ }
                                                                                                    & \multicolumn{2}{c}{$\Delta g_{\rm SE}$ (ext.)} \\
\colrule
\endhead
\hline \hline
\endfoot
1  &     1.&52928  &   2320.&77563    &    0.&50250    &     0.&03305   &
2322.&84046(10)  &    0.&00000  &  2322.&84046(10)       \\
   &       &       &        &         &      &         &       &        &        &           &      &       &  2322.&8404(9)$^a$    \\
2  &     5.&20640  &   2316.&00970    &    1.&55757    &     0.&13053   &
2322.&90420(9)   &    0.&00000  &  2322.&90420(9)        \\
   &       &       &        &         &      &         &       &        &        &           &      &       &  2322.&9040(9)$^a$    \\
3  &    10.&52313  &   2309.&28506    &    2.&91759    &     0.&28869   &
2323.&01447(9)   &    0.&00000  &  2323.&01447(9)        \\
   &       &       &        &         &      &         &       &        &        &           &      &       &  2323.&0140(9)$^a$    \\
4  &    17.&21613  &   2300.&99753    &    4.&45945    &     0.&50260   &
2323.&17571(9)   &    0.&00000  &  2323.&17571(9)        \\
   &       &       &        &         &      &         &       &        &        &           &      &       &  2323.&1751(9)$^a$    \\
5  &    25.&10744  &   2291.&41521    &    6.&10392    &     0.&76661   &
2323.&39318(9)   &    0.&00000  &  2323.&39318(9)        \\
   &       &       &        &         &      &         &       &        &   2323.&42(5)$^b$  &      &       &  2323.&3928(9)$^a$    \\
6  &    34.&06467  &   2280.&73799    &    7.&79535    &     1.&07460   &
2323.&67261(9)   &    0.&00000  &  2323.&67261(9)        \\
   &       &       &        &         &      &         &       &        &        &           &      &       &  2323.&6724(9)$^a$    \\
8  &    54.&78171  &   2256.&69788    &   11.&16571    &     1.&79701   &
2324.&44230(9)   &   -0.&00001  &  2324.&44229(9)        \\
   &       &       &        &         &      &         &       &        &        &           &      &       &  2324.&4421(10)$^a$   \\
10 &    78.&74380  &   2229.&82629    &   14.&34905    &     2.&61754   &
2325.&53668(10)  &   -0.&00002  &  2325.&53666(10)       \\
   &       &       &        &         &      &         &       &        &   2325.&28$^b$     &      &       &  2325.&5355(10)$^a$   \\
12 &   105.&51169  &   2200.&79830    &   17.&21623    &     3.&48376   &
2327.&00998(12)  &   -0.&00005  &  2327.&00993(12)       \\
   &       &       &        &         &      &         &       &        &        &           &      &       &  2327.&0103(12)$^a$   \\
15 &   150.&23524  &   2154.&28732    &   20.&77184    &     4.&75746   &
2330.&05186(16)  &   -0.&00011  &  2330.&05175(16)       \\
   &       &       &        &         &      &         &       &        &   2329.&79$^b$     &      &       &  2330.&051(1)$^a$     \\
18 &   199.&76448  &   2105.&29703    &   23.&34254    &     5.&85654   &
2334.&26059(20)  &   -0.&00029  &  2334.&26030(20)       \\
   &       &       &        &         &      &         &       &        &        &           &      &       &  2334.&262(2)$^a$     \\
20 &   235.&17645  &   2071.&71455    &   24.&49971    &     6.&41815   &
2337.&80885(24)  &   -0.&00052  &  2337.&80833(24)       \\
   &       &       &        &         &      &         &       &        &   2337.&50$^b$     &      &       &  2337.&86(1)$^a$      \\
24 &   311.&21758  &   2003.&24295    &   25.&53187    &     6.&95150   &
2346.&9439(3)    &   -0.&0015   &  2346.&9424(3)         \\
   &       &       &        &         &      &         &       &        &        &           &      &       &  2346.&92(1)$^a$      \\
30 &   437.&19791  &   1899.&42146    &   24.&22809    &     5.&76765   &
2366.&6151(3)    &   -0.&0057   &  2366.&6094(3)         \\
   &       &       &        &         &      &         &       &        &   2366.&77$^b$     &      &       &  2366.&59(1)$^a$      \\
32 &   482.&18213  &   1864.&94188    &   23.&14976    &     4.&73024   &
2375.&0040(4)    &   -0.&0088   &  2374.&9952(4)         \\
   &       &       &        &         &      &         &       &        &        &           &      &       &  2374.&97(1)$^a$      \\
40 &   676.&59132  &   1729.&56158    &   16.&47453    &  $-$3.&19027   &
2419.&4372(5)    &   -0.&0349   &  2419.&4023(5)         \\
   &       &       &        &         &      &         &       &        &   2419.&45$^b$     &      &       &  2419.&39(1)$^a$      \\
50 &   952.&87040  &   1569.&36938    &    5.&18971    & $-$22.&43986   &
2504.&9896(7)    &   -0.&1615   &  2504.&8281(7)         \\
   &       &       &        &         &      &         &       &        &   2504.&09$^b$     &      &       &  2504.&827(8)$^a$     \\
54 &  1074.&52868  &   1508.&85111    &    0.&51786    & $-$33.&12979   &
2550.&768(2)     &   -0.&282    &  2550.&486(2)          \\
   &       &       &        &         &      &         &       &        &        &           &      &       &  2550.&487(8)$^a$     \\
60 &  1270.&36538  &   1422.&34049    & $-$6.&00306    & $-$52.&2051    &
2634.&498(3)     &   -0.&610    &  2633.&888(3)          \\
   &       &       &        &         &      &         &       &        &   2634.&54$^b$     &      &       &  2633.&895(9)$^a$     \\
70 &  1638.&25837  &   1290.&32234    &$-$14.&07335    & $-$90.&9411    &
2823.&566(5)     &   -2.&193    &  2821.&373(5)          \\
   &       &       &        &         &      &         &       &        &   2823.&39$^b$     &      &       &  2821.&39(1)$^a$      \\
80 &  2072.&57316  &   1174.&47096    &$-$16.&61904    &$-$135.&1054    &
3095.&320(10)    &   -6.&913    &  3088.&407(10)         \\
   &       &       &        &         &      &         &       &        &   3095.&34$^b$     &      &       &  3088.&46(2)$^a$      \\
82 &  2169.&55512  &   1153.&32686    &$-$16.&29951    &$-$144.&1073    &
3162.&475(12)    &   -8.&687    &  3153.&788(12)         \\
   &       &       &        &         &      &         &       &        &        &           &      &       &  3153.&85(2)$^a$      \\
90 &  2601.&21377  &   1075.&78727    &$-$11.&90297    &$-$178.&5734    &
3486.&525(20)    &  -22.&32     &  3464.&205(20)         \\
   &       &       &        &         &      &         &       &        &   3486.&56(3)$^a$  &      &       &  3464.&35(2)$^a$      \\
   &       &       &        &         &      &         &       &        &   3487.&30$^b$     &      &       &       &                \\
92 &  2722.&17025  &   1058.&21204    & $-$9.&99010    &$-$186.&3945    &
3583.&998(20)    &  -28.&12     &  3555.&878(20)         \\
   &       &       &        &         &      &         &       &        &        &           &      &       &  3556.&05(2)$^a$      \\

\end{longtable*}
$^a$ Ref. \cite{Beier00}, $^b$ Ref. \cite{Blundell97}.
\endgroup

%
%
\begingroup
\begin{longtable*}{rr@{}lr@{}lr@{}lr@{}lr@{}lr@{}lr@{}l}
\caption{ Various contributions to the one-loop self-energy correction to the
$2s$-electron $g$-factor for H-like ions for the point nuclear model. All
values are absolute contributions to the $g$-factor ($1/\alpha =
137.035\,989\,5$) and presented in units of $10^{-6}$ (ppm). \label{table2s}}\\
\hline
\hline
$Z$ & \multicolumn{2}{c}{$\Delta g_{\rm ir}$} &
                    \multicolumn{2}{c}{$\Delta g_{\rm vr}^{(0)}$} &
                                    \multicolumn{2}{c}{$\Delta g_{\rm vr}^{(1)}$}
                                                   & \multicolumn{2}{c}{$\Delta g_{\rm vr}^{(2+)}$}
                                                                &\multicolumn{2}{c}{ $\Delta g_{\rm SE}$ (pnt.)}\\
\colrule
\endfirsthead
\caption{$2s$ $g$-factor. (continued)}\\
\hline
\hline
$Z$ & \multicolumn{2}{c}{$\Delta g_{\rm ir}$} &
                    \multicolumn{2}{c}{$\Delta g_{\rm vr}^{(0)}$} &
                                    \multicolumn{2}{c}{$\Delta g_{\rm vr}^{(1)}$}
                                                   & \multicolumn{2}{c}{$\Delta g_{\rm vr}^{(2+)}$}
                                                                &\multicolumn{2}{c}{ $\Delta g_{\rm SE}$ (pnt.)}\\
\colrule
\endhead
\hline
\hline
\endfoot
2  &    1.&4905   &   2320.&7711   &      0.&4471    &     0.&1317  &  2322.&8404(3)  \\
4  &    5.&0539   &   2315.&9889   &      1.&3379    &     0.&5245  &  2322.&9051(4)   \\
6  &   10.&1863   &   2309.&2305   &      2.&4286    &     1.&1729  &  2323.&0183(6)  \\
8  &   16.&629    &   2300.&886    &      3.&600     &     2.&070   &  2323.&185(1)    \\
10 &   24.&209    &   2291.&216    &      4.&778     &     3.&210   &  2323.&413(2)   \\
12 &   32.&796    &   2280.&417    &      5.&909     &     4.&585   &  2323.&707(2)    \\
14 &   42.&290    &   2268.&639    &      6.&956     &     6.&188   &  2324.&074(3)   \\
16 &   52.&609    &   2256.&005    &      7.&892     &     8.&014   &  2324.&520(3)    \\
18 &   63.&683    &   2242.&618    &      8.&696     &    10.&056   &  2325.&052(5)   \\
20 &   75.&453    &   2228.&563    &      9.&351     &    12.&307   &  2325.&674(5)    \\
24 &  100.&887    &   2198.&735    &     10.&177     &    17.&427   &  2327.&225(5)   \\
30 &  143.&170    &   2150.&542    &     10.&109     &    26.&584   &  2330.&405(5)    \\
32 &  158.&234    &   2133.&739    &      9.&728     &    30.&024   &  2331.&726(6)   \\
40 &  222.&652    &   2063.&747    &      6.&429     &    45.&708   &  2338.&536(8)    \\
50 &  311.&098    &   1971.&870    &   $-$1.&448     &    69.&823   &  2351.&343(9)   \\
54 &  348.&579    &   1934.&243    &   $-$5.&641     &    81.&003   &  2358.&184(9)    \\
60 &  406.&838    &   1877.&219    &  $-$12.&891     &    99.&640   &  2370.&807(9)   \\
70 &  509.&241    &   1781.&353    &  $-$27.&042     &   136.&596   &  2400.&149(9)    \\
80 &  619.&038    &   1685.&423    &  $-$42.&849     &   183.&153   &  2444.&765(9)   \\
82 &  642.&093    &   1666.&316    &  $-$46.&101     &   193.&936   &  2456.&245(9)    \\
83 &  653.&785    &   1656.&779    &  $-$47.&729     &   199.&543   &  2462.&378(9)   \\
90 &  739.&302    &   1590.&411    &  $-$59.&021     &   243.&372   &  2514.&064(9)    \\
92 &  765.&177    &   1571.&607    &  $-$62.&163     &   257.&585   &  2532.&207(9)    \\
\end{longtable*}
\endgroup

\end{document}